%% file: main.tex
\documentclass[letterpaper]{article} % DO NOT CHANGE THIS
\usepackage{aaai25}  % Camera-ready: submission (anonymous) mode removed
\usepackage{times}  % DO NOT CHANGE THIS
\usepackage{helvet}  % DO NOT CHANGE THIS
\usepackage{courier}  % DO NOT CHANGE THIS
\usepackage[hyphens]{url}  % DO NOT CHANGE THIS
\usepackage{graphicx} % DO NOT CHANGE THIS
\usepackage{natbib}  % DO NOT CHANGE THIS AND DO NOT ADD ANY OPTIONS TO IT
\usepackage{caption} % DO NOT CHANGE THIS AND DO NOT ADD ANY OPTIONS TO IT
\usepackage{amsmath}
\usepackage{booktabs}
\usepackage{seqsplit}
\usepackage{enumitem}
\usepackage{comment}
\providecommand{\pdfinfo}[1]{} % no-op fallback: not a primitive under this engine outside submission mode
\input{tables/macros}

\input{tables/macros_ext}

\input{tables/macros_media}
\input{tables/macros_media_robustness}

\input{tables/macros_dailyleadlag}


\input{tables/macros_sample}

\input{tables/macros_disappear}
\input{tables/macros_disappear_reverse}
\input{tables/macros_scores}

\input{tables/macros_scorehist}
\input{tables/macros_evbehavior}
\input{tables/macros_behavior}


\input{tables/macros_suspension}

\input{tables/macros_suspension_adjusted}

\input{tables/macros_activity}

\input{tables/macros_eventstatus}
\input{tables/macros_content}

\input{tables/macros_filterdiag}


\input{tables/macros_shares}

\input{tables/macros_spikefreq}

\input{tables/macros_events_checking}

\input{tables/macros_theme_table}





\input{tables/macros_ncheck}



\input{tables/macros_dispersion}
\input{tables/macros_rq2_consistent}
\input{tables/macros_mentionpersist}













\usepackage{xcolor}
\usepackage{soul}
\usepackage{placeins}
\usepackage{float}

\newcommand{\answerYes}[1]{\textcolor{blue}{#1}}
\newcommand{\answerNo}[1]{\textcolor{teal}{#1}}
\newcommand{\answerNA}[1]{\textcolor{gray}{#1}}

\title{When Does the Public Become Suspicious of Bots?\\
Demand-Side Evidence from Botometer Query Logs}
\author{Tuğrulcan Elmas}
\affiliations{Observatory on Social Media, Indiana University, Bloomington, USA \\ University of Edinburgh, Edinburgh, UK}
\nocopyright

\begin{document}
\maketitle

\input{sections/abstract}

\input{sections/introduction}

\input{sections/related_work}

\input{sections/data}

\input{sections/methods}

\input{sections/results}

\input{sections/limitations}

\input{sections/acknowledgements}

{
\small
\bibliography{references}
}

\input{sections/checklist}

\input{sections/ethics}

\appendix
\raggedbottom
% Superseded/merged/relocated appendices and their reorganization history:
% legacy/main_tex_appendix_history.txt
\input{sections/appendix_robustness}

\input{sections/appendix_placebo}

\input{sections/appendix_k}

\input{sections/appendix_disappear_activity}

\end{document}

%% file: tables/macros.tex
\newcommand{\nQueries}{250,838,846}
\newcommand{\dateMin}{2020-09-01}
\newcommand{\dateMax}{2023-06-30}
\newcommand{\nTargets}{99,225,402}
\newcommand{\nIps}{219,414}
\newcommand{\nBulkIps}{4,280}
\newcommand{\bulkShare}{97.5}
\newcommand{\nCadenceIps}{76,201}
\newcommand{\cadenceShare}{77.9}
\newcommand{\nOrganic}{1,379,541}
\newcommand{\nOrganicIps}{138,933}
\newcommand{\nOrganicTargets}{834,749}
\newcommand{\requestersPerDay}{218}
\newcommand{\nChecks}{1,182,588}

\newcommand{\inflationShare}{14.3}
\newcommand{\pctOnce}{85.3}
\newcommand{\topHundredShare}{2.8}
\newcommand{\topThousandShare}{7.2}
\newcommand{\topTenThousandShare}{15.6}
\newcommand{\nEvents}{1,859}

\newcommand{\pctDefaultHandle}{9.6}

%% file: tables/macros_ext.tex
\newcommand{\habitPctOneDay}{82.9}
\newcommand{\habitPctReturn}{17.1}
\newcommand{\habitPctTenDays}{1.1}

\newcommand{\habitReturnCheckShare}{58.4}

%% file: tables/macros_media.tex
\newcommand{\mediaEventsTwitterBotsR}{0.48}
\newcommand{\mediaEventsTwitterBotsT}{3.1}
\newcommand{\mediaEventsTwitterBotsRho}{0.50}
\newcommand{\mediaChecksTwitterBotsR}{0.41}
\newcommand{\mediaChecksTwitterBotsT}{2.6}
\newcommand{\mediaChecksTwitterBotsRho}{0.60}

\newcommand{\mediaEventsBotometerR}{0.41}
\newcommand{\mediaEventsBotometerT}{2.5}
\newcommand{\mediaEventsBotometerRho}{0.28}
\newcommand{\mediaChecksBotometerR}{0.30}
\newcommand{\mediaChecksBotometerT}{1.8}
\newcommand{\mediaChecksBotometerRho}{0.27}

\newcommand{\mediaEventsRedditR}{0.26}
\newcommand{\mediaEventsRedditT}{1.5}
\newcommand{\mediaEventsRedditRho}{0.23}
\newcommand{\mediaChecksRedditR}{0.33}
\newcommand{\mediaChecksRedditT}{2.0}
\newcommand{\mediaChecksRedditRho}{0.42}

\newcommand{\mediaEventsMuskR}{0.15} % [preserved]
\newcommand{\mediaEventsMuskT}{0.9} % [preserved]
\newcommand{\mediaEventsMuskRho}{0.08} % [preserved]
\newcommand{\mediaChecksMuskR}{0.11} % [preserved]
\newcommand{\mediaChecksMuskT}{0.6} % [preserved]
\newcommand{\mediaChecksMuskRho}{-0.04} % [preserved]

%% file: tables/macros_media_robustness.tex
\newcommand{\mediaEventsTwitterBotsDiffR}{0.03}

\newcommand{\mediaChecksTwitterBotsDiffR}{-0.04}

\newcommand{\mediaEventsBotometerDiffR}{0.22}

\newcommand{\mediaChecksBotometerDiffR}{-0.09}

%% file: tables/macros_dailyleadlag.tex
\newcommand{\dailyChecksRedditR}{-0.02}

\newcommand{\dailyDiffRedditR}{-0.08}

%% file: tables/macros_sample.tex
\newcommand{\bgHandleTweet}{7.7}
\newcommand{\bgHandleUser}{8.8}
\newcommand{\bgMedianAge}{728}
\newcommand{\bgPctYoung}{9.2}

\newcommand{\targetCoverage}{82.3}
\newcommand{\targetMatched}{687,346}

\newcommand{\targetMedianAgeExact}{1,418}
\newcommand{\targetPctYoungExact}{4.4}

\newcommand{\langEnChecked}{61.8}
\newcommand{\langEnBg}{33.7}
\newcommand{\langJaChecked}{1.4}
\newcommand{\langJaBg}{18.6}

\newcommand{\archiveTweets}{3.2 billion}

\newcommand{\evAccountsN}{736}
\newcommand{\evElevMean}{5.5}

\newcommand{\evElevLeOnePct}{57.3}
\newcommand{\evElevMidPct}{16.7}
\newcommand{\evElevGeTwoPct}{26}

%% file: tables/macros_disappear.tex
\newcommand{\disTreatedPct}{17.1}
\newcommand{\disControlPct}{14.3}
\newcommand{\disRatio}{1.20}
\newcommand{\disTreatedN}{367,244}

\newcommand{\disZ}{43.1}
\newcommand{\disControlAdjPct}{12.3}

%% file: tables/macros_disappear_reverse.tex
\newcommand{\disTreatedRevStdPct}{18.8}
\newcommand{\disRatioRev}{1.31}

%% file: tables/macros_scores.tex
\newcommand{\scoreN}{834,749}
\newcommand{\scoreMedian}{0.35}
\newcommand{\scoreHighPct}{16.5}
\newcommand{\scoreLowPct}{31.5}
\newcommand{\scoreEventMedian}{0.63}
\newcommand{\scoreEventHighPct}{28.3}

\newcommand{\scoreGoneQOne}{13.5}
\newcommand{\scoreGoneQFour}{19.8}
\newcommand{\scoreQOneMax}{0.13}
\newcommand{\scoreQFourMin}{0.58}

%% file: tables/macros_scorehist.tex
\newcommand{\scoreDisRateMin}{13.2}
\newcommand{\scoreDisRateMax}{22.2}
\newcommand{\scoreSuspRateMin}{6.5}
\newcommand{\scoreSuspRateMax}{23.3}

%% file: tables/macros_evbehavior.tex
\newcommand{\evbActiveElevatedPct}{39.1}
\newcommand{\evbBurstOnlyPct}{6.8}

\newcommand{\evbMostlyRetweetsPct}{69.2}

\newcommand{\evbSpamPct}{3.6}

\newcommand{\evbActiveUncapturedPct}{41.8} % [preserved]
\newcommand{\evbDormantPct}{5.6} % [preserved]
\newcommand{\evbLowActivityPct}{8.6} % [preserved]

%% file: tables/macros_behavior.tex
\newcommand{\burstObsElevatedPct}{16.2}
\newcommand{\burstPlaceboElevatedPct}{10.9}

\newcommand{\burstActivePct}{23.9}
\newcommand{\burstPlaceboActivePct}{16.5}

\newcommand{\burstPostOneActivePct}{22.5}
\newcommand{\burstPostOneElevatedPct}{14.7}

\newcommand{\burstPostWeekActivePct}{21.0}
\newcommand{\burstPostWeekElevatedPct}{13.3}

\newcommand{\behScoredN}{563,111}

%% file: tables/macros_suspension.tex
\newcommand{\usSampleN}{5,000}
\newcommand{\usGoneN}{2,500}
\newcommand{\usPresentN}{2,500}

\newcommand{\usGoneSuspPct}{11.0}
\newcommand{\usPresentSuspPct}{13.3}

\newcommand{\usSuspP}{0.014}

%% file: tables/macros_suspension_adjusted.tex
\newcommand{\susAdjOR}{2.28}
\newcommand{\susAdjORLow}{1.79}
\newcommand{\susAdjORHigh}{2.91}

\newcommand{\susAdjZ}{6.7}
\newcommand{\susAdjP}{<0.001}
\newcommand{\susRawOR}{2.31}
\newcommand{\susRawORLow}{1.90}
\newcommand{\susRawORHigh}{2.82}

%% file: tables/macros_activity.tex
\newcommand{\goneAliveActN}{1,742}

\newcommand{\goneAliveUnavailPct}{4.4}
\newcommand{\goneAliveActWithTweetsN}{1,623}
\newcommand{\goneAliveActNoVisiblePct}{2.6}
\newcommand{\goneAliveDormantYearOnePct}{28.6}
\newcommand{\goneAliveDormantYearTwoPct}{39.2}
\newcommand{\goneAliveDormantYearThreePct}{50.2}

\newcommand{\eventDormantYearOnePct}{27.2}

\newcommand{\presentActWithTweetsN}{1,683}

\newcommand{\presentDormantYearOnePct}{9.8}
\newcommand{\presentDormantYearTwoPct}{17.9}
\newcommand{\presentDormantYearThreePct}{26.4}

%% file: tables/macros_eventstatus.tex
\newcommand{\evStatusSuspPct}{22.3}
\newcommand{\evStatusGonePct}{18.9}
\newcommand{\baseSuspPct}{12.1}
\newcommand{\baseGonePct}{17.2}
\newcommand{\evStatusSuspZ}{7.8}
\newcommand{\evStatusSuspP}{<0.001}
\newcommand{\evStatusGoneZ}{1.1}
\newcommand{\evStatusGoneP}{0.265}
\newcommand{\baseSuspReweightedPct}{12.9}
\newcommand{\evStatusSuspReweightedZ}{6.0}
\newcommand{\evStatusSuspReweightedP}{<0.001}

%% file: tables/macros_content.tex
\newcommand{\ctBioEmpty}{14.2}

\newcommand{\ctRtShare}{39.9}

\newcommand{\ctUrlShare}{29.4}
\newcommand{\ctTagShare}{15.4}
\newcommand{\ctMentionShare}{65.9}

\newcommand{\cbBioEmpty}{17.8}

\newcommand{\cbRtShare}{44.5}

\newcommand{\cbUrlShare}{21.4}
\newcommand{\cbTagShare}{7.6}
\newcommand{\cbMentionShare}{53.0}

\newcommand{\ratioBioEmpty}{0.8}

%% file: tables/macros_filterdiag.tex
\newcommand{\fdVolAlsoCadPct}{90}

%% file: tables/macros_shares.tex
\newcommand{\shOrganicTargets}{0.8}
\newcommand{\shEventAccounts}{801}

%% file: tables/macros_spikefreq.tex
\newcommand{\evSpikeOncePct}{77.7}
\newcommand{\evSpikeMidPct}{18.4}
\newcommand{\evSpikeTailPct}{4.0}
\newcommand{\evSpikeMaxCount}{133}

%% file: tables/macros_events_checking.tex
\newcommand{\evChecksEventsR}{0.77}
\newcommand{\evChecksEventsRho}{0.63}
\newcommand{\evChecksEventsN}{34}
\newcommand{\evPeakChecksGrowth}{2.8}
\newcommand{\evPeakEventsGrowth}{6.0}
\newcommand{\evPreBreakEventsMean}{21}
\newcommand{\evPostBreakEventsMean}{75}
\newcommand{\evPrePostEventsGrowth}{3.6}
\newcommand{\evPreBreakChecksMean}{30,723}
\newcommand{\evPostBreakChecksMean}{37,270}

%% file: tables/macros_theme_table.tex
\newcommand{\fbThemeAllN}{5,710}
\newcommand{\fbThemeBotN}{1,285}
\newcommand{\fbThemeCyborgN}{897}
\newcommand{\fbThemeHumanN}{3,528}

%% file: tables/macros_ncheck.tex
\newcommand{\ncN}{834,749}
\newcommand{\ncRho}{0.05}
\newcommand{\ncMeanRho}{0.05}
\newcommand{\ncOnceMed}{0.34}
\newcommand{\ncTopLabel}{50+}
\newcommand{\ncTopMed}{0.62}
\newcommand{\ncTopN}{530}

%% file: tables/macros_dispersion.tex
\newcommand{\targetMedianStatuses}{6,461}
\newcommand{\bgMedianStatuses}{3,993}
\newcommand{\statusesRatio}{1.6}
\newcommand{\onceMedFollowers}{495}
\newcommand{\multiMedFollowers}{2,608}
\newcommand{\onceMedStatuses}{5,717}
\newcommand{\multiMedStatuses}{13,319}
\newcommand{\onceMedAge}{1,435}
\newcommand{\multiMedAge}{1,394}
\newcommand{\tsHtBgMed}{0.00}
\newcommand{\tsHtOnceMed}{0.02}
\newcommand{\tsHtMultiMed}{0.08}
\newcommand{\tsLenBgMed}{49}
\newcommand{\tsLenOnceMed}{96}
\newcommand{\tsLenMultiMed}{109}
\newcommand{\tsUrlBgMed}{0.00}
\newcommand{\tsUrlOnceMed}{0.17}
\newcommand{\tsUrlMultiMed}{0.27}
\newcommand{\tsBgHashtagZeroPct}{87.9}
\newcommand{\tsBgUrlZeroPct}{75.3}

%% file: tables/macros_rq2_consistent.tex
\newcommand{\bgCreationMed}{2019-09-15}
\newcommand{\targetCreationMed}{2018-01-24}

%% file: tables/macros_mentionpersist.tex
\newcommand{\evElevPostOneMean}{3.3}
\newcommand{\evElevPostWeekMean}{1.6}

%% file: sections/abstract.tex
\begin{abstract}
We study private bot-checking behavior from the
\emph{demand} side: when people suspect an account is
automated, whom they suspect, and what follows. Using Botometer's
server-side query logs, the most widely used bot-detection service,
we treat each query as a behavioural trace of suspicion. We analyze over 1 million public checks of Twitter accounts
from 2020 to 2023, enriched with \archiveTweets{} tweets of
the contemporaneous 1\% public stream. Collective suspicion spikes with platform crises, most sharply around the 2022 Musk--Twitter bot dispute. Checked
accounts are older and more prolific, have more followers, and post promotional, political, and crypto content. Accounts that draw collective suspicion have higher bot scores and are more likely to be suspended. Bot-related public attention and Botometer activity are elevated during the same broad periods, although their short-run fluctuations are largely uncoupled. Public feedback focuses on first-person identity claims for \emph{human}, and evidence-based arguments citing posting rate, political content, and
cross-account coordination for \emph{bot}/\emph{cyborg}. Bot suspicion thus constitutes a mass, distributed form of platform auditing that tracks meaningful signals of automation, establishing audit-tool query logs as a novel lens on public responses to platform manipulation.

\end{abstract}

%% file: sections/introduction.tex
\section{Introduction}

Social bots research has produced detection systems,
prevalence estimates, and rich characterisations of automated behaviour
\cite{cresci2020decade}. The literature primarily focuses on bots from
the supply side: it asks what type of bots exist and what they do. Yet the
social significance of bots depends equally on the demand side, on the public that worries about them. Fears of automated manipulation shape trust in online discourse, figure prominently in high-profile disputes over platform governance, as when bot-prevalence estimates became Elon Musk's stated grounds for attempting to exit the 2022 Twitter acquisition deal \citep{bond2022botsclaim}, and motivate the accusation ``you're a bot,'' a staple of online argument. Recent developments raise the salience of that worry further. Platforms such as X inadvertently provided incentives to bot farms through an ad revenue sharing program that encourages engagement~\citep{molla2024xbots}.  Large language models have simultaneously
lowered the cost of producing convincing, human-sounding bot
accounts to near zero \citep{yangmenczer2024botnet}. Financial incentives and low costs made engagement-maximising ``hook'' and rage-bait content prevalent~\citep{cover2025ragebait}.
These issues make automated, engagement-optimised accounts both cheaper to run and more directly profitable than before. This shift is a plausible driver of the rising rate of public bot accusations documented in prior work \citep{assenmacher2023bot}. Meanwhile, there is limited empirical evidence regarding the conditions under which individuals become suspicious of bot activity and which users are most likely to be perceived as bots.
The obstacle has been data: suspicion is a private cognitive state that
rarely leaves a trace. To overcome this obstacle, we employ Botometer's internal dataset.
Botometer
\cite{davis2016botornot,yang2022botometer} is a public web service that
scores Twitter / X accounts on their likelihood of being a bot, i.e., a partially or a fully automated account. Thousands of users, including researchers, journalists, and curious members of the public, submit accounts to be checked. We treat each query as a behavioural trace of suspicion: a check indicates that an account was salient enough, or uncertain enough, for someone to seek a bot assessment, although the query itself does not reveal the user's underlying motive. The
service's server-side logs therefore constitute, to our knowledge, the first large-scale observational dataset of \textit{public bot-checking behavior} in the wild. We ask three questions:
\begin{enumerate}[label=RQ\arabic*:, leftmargin=*, itemsep=4pt, topsep=4pt]
\item What are the characteristics of suspicious accounts checked by the public?

\item When does suspicion become collective?

\item What comes of suspicion?
\end{enumerate}

Our contributions are: (1) a longitudinal and large-scale
characterisation of public bot suspicion, covering its volume, rhythm and burst
structure, (2) a characterisation
of the targets of suspicion, benchmarked against \archiveTweets{}
tweets of the contemporaneous 1\% stream, showing that suspicion is a
mass practice aimed principally at ordinary, established, political
English-language accounts rather than at elites, newly created
accounts, or fans, and
(3) the first linkage of
public suspicion to its outcomes, showing that checked accounts subsequently
cease visible activity more often than comparable accounts.

%% file: sections/related_work.tex
\section{Related Work}

\noindent\textbf{Supply-side bot research.}
Most work on social bots concerns their detection and characterisation:
feature-based classifiers
\cite{davis2016botornot,yang2023challenges}, censuses of bot behaviour
and influence \cite{varol2017online}, bots' humanlike behaviours \cite{cresci2020decade}, and their coordinated activity~\cite{pote2025coordinated}.
Multi-platform ensemble detectors are applied to specific events such as
the 2020 U.S. election \cite{ng2024multiplatform}, and network-based coordination detection applied to conflict related discussions \cite{wuyu2026russiaukraine, elmas2026israelhamas}, and trending topics~\cite{gopalakrishnan2025density,gopalakrishnan2025engagement}. Detection has also moved beyond
Twitter/X and beyond single accounts acting alone: to
messaging-platform bots at scale \cite{tsuchiya2026telegram}, to
network-level detection of coordinated inauthentic behaviour on
video-first platforms \cite{luceri2026coordinated}, and to
behavioural \emph{change} over an account's
lifetime as a signal that generalises across both automation and
coordination \cite{ariyarathne2026behavior}. The arrival of capable
language models has pushed this literature to ask which features
survive an adversary who can generate human-sounding text at
near-zero cost. One proposed answer is account-history features, which are
harder to fake than post content
\cite{katyal2026accounthistory}. The literature asks what bots are
and do. We study bot suspicion itself, from a demand-side, consumer-oriented
perspective, using data collected through a public tool.

\noindent\textbf{Botometer as a public instrument.}
Botometer was explicitly framed as ``arming the public'' against
automation \cite{yang2019arming}, and its operators note that the free website is aimed at lay users
\cite{yang2022botometer}. It is one instance of a broader class: public-facing tools that
let ordinary users audit an opaque automated system rather than leaving
that oversight to platforms or regulators. The algorithmic-auditing literature distinguishes such \emph{end-user audits} from formal, third-party audits and has begun designing tools to support them, for instance for personalized recommendation feeds \cite{wu2024mapmyfeed} and for investigating why an account became popular in the first place \cite{elmas2022waypop}. 
In parallel, a broader governance literature examines ethics-based auditing of automated decision systems \cite{mokander2021ethicsbased}. Researcher-run audits of the same
opaque-system kind have examined how platforms curate what users see,
such as Reddit's algorithmically ranked r/popular feed
\cite{chan2026curation}, sock-puppet audits of comment ranking on
TikTok \cite{yan2026auditing}, and an audit of AI-generated-text
detectors, which similarly turns a detection tool into an object of
study rather than an instrument \cite{dutta2026audit}. This body of work either asks how to make end-user audits more
effective or, like the AI-generated-text study above, turns the audit
tools themselves into an object of scrutiny. By contrast,
our query log lets us observe one such tool's actual,
unprompted use at scale.

\noindent\textbf{The demand side: suspecting and accusing.}
Closest to our work, \citet{assenmacher2023bot} study \emph{bot accusations} on Twitter, public replies that label an account a bot, and find that such accusations increasingly serve to discredit and dehumanize others in political discussions. By contrast, our query logs capture the complementary,
\emph{private} act of verification, invisible to the accused and to
other users. They are paired with the moment-of-suspicion timestamp, facilitating temporal analysis.
Survey and experimental work shows humans are poor at identifying
automated accounts \cite{radivojevic2024llms}, and diffuse anxiety
about an automated web has crystallized into the ``dead internet theory'' \cite{muzumdar2025dead}. Closest to our finding that accounts drawing suspicion from many independent checkers score more reliably than those checked only once, a study of collaborative AI-content labelling on Reddit similarly finds that aggregation acts as a reliability filter on individually noisy judgements \cite{elmas2026communities}.
A related literature uses surveys and interviews to examine users' \emph{attitudes} towards platform content moderation. Users express less support for moderation as interventions become more aggressive, from labelling to content removal \cite{urman2023contentmoderation}. They also prefer greater control over what content is moderated for them and who performs that moderation \cite{jhaver2023personalizing}. Closer to automation specifically, an interactive simulation of AI-driven profiling finds that even privacy-conscious users can be accurately profiled, and that experiencing such profiling changes their stated willingness to disclose information in the future \cite{yu2026neighborhood}. Similarly, a labeling experiment finds that marking
content as AI-generated lowers its own perceived authenticity, while increasing the perceived authenticity of nearby unlabelled content by contrast
\cite{pawelczyk2026implied}. These studies ask what people think about moderation, automation, or authenticity through surveys, interviews, or controlled exposure. In contrast, our query logs capture a moderation-adjacent behaviour directly through the unprompted act of checking. Capturing such natural behaviour makes our study less susceptible to observer effects. We provide the first large-scale behavioural measurement of this kind.

%% file: sections/data.tex
\section{Data}

Botometer is a public,
free-to-use web service, developed and maintained by the Indiana
University Observatory on Social Media, that scores a Twitter/X
account's likelihood of being automated. A user submits an account
by screen name or, through the API, in bulk. The service pulls
that account's recent activity and network, extracts on the order of
a thousand features spanning content, sentiment, network structure,
and temporal posting patterns, and returns a score in $[0,1]$ from a
supervised ensemble classifier, where higher values indicate a more
bot-like profile. There are two score variants: an \emph{English} score
trained on English-language content features and a
\emph{universal} score that omits language-dependent features and so
applies across languages. No fixed
threshold is codified as ``the'' bot cutoff, but scores near 0 or 1 are conventionally read as confidently human-like or bot-like, and we follow this convention when we report thresholds at 0.2 and 0.8.
Botometer has operated continuously since 2014 and is used both by
researchers, who query it programmatically over account lists, and by
the public who query the accounts they
personally encounter and find suspicious. The latter is our primary focus, and the query log is the only source that lets us observe it.

\noindent\textbf{Botometer query log:}
Our primary dataset is the server-side query log of the Botometer web interface and
API. Each record contains the queried (target) account's numeric user
ID and screen name, a timestamp, the requester's id, and
the bot scores the service returned (English and universal, in
$[0,1]$), and the target's most recent tweet timestamp and daily tweeting rate. The log spans \dateMin{} to \dateMax{} and contains
\nQueries{} queries to \nTargets{} unique accounts, spanning crucial events such as the 2020 U.S. election, the
January 6, 2021 attack on the U.S. Capitol, and the 2022 Musk acquisition of
Twitter, from his April bid to the deal's close that
October~\cite{lerman2022muskbid,perez2022muskbuystwitter}. Checking reaches its
single-month record of 79{,}621 checks in January 2023, then declines as Twitter tightens API access over the following months,
collapsing from roughly 520 checks/day in June to zero by early
July. We take June 30, 2023, the
last day of sustained operation, as the end of the analysis window. The tool was revived as Botometer~X in September 2024 to provide past bot scores. We exclude it from the analysis.

 \begin{figure}[t]
\centering
\includegraphics[width=\columnwidth]{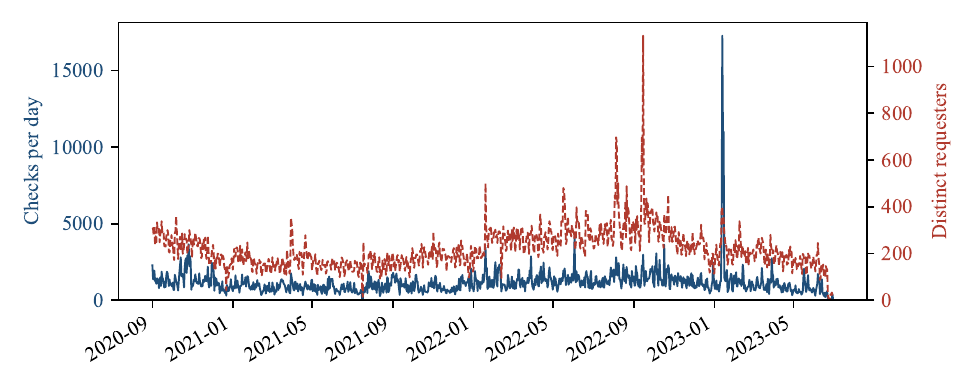}
\caption{Daily checks (solid) \& distinct requesters
(dashed)}
\label{fig:daily}
\end{figure}

\noindent\textbf{Feedback log:}
After a lookup, Botometer's public interface lets users label an account as bot, human, or cyborg, and optionally explain their choice. The log contains \fbThemeAllN{} responses (\fbThemeBotN{} bot, \fbThemeCyborgN{} cyborg, \fbThemeHumanN{} human).

\noindent\textbf{Stream sample:}
We employ the Internet Archive's Twitter Stream Grab
\cite{archiveteam2023twitterstream}, a
continuous $\sim$1\% sample of the public tweet stream, which covers our window up to January 2023 and provides
\archiveTweets{} tweets. It complements the log with tweet language and account creation date, a
baseline population of accounts \emph{not} checked, an independent
measure of each target's conversational visibility, and a behavioural
record of whether targets remain active after being checked. As the stream samples accounts in proportion to their tweeting
activity, the matched targets over-represent active accounts, which we acknowledge in Limitations.

\noindent\textbf{External attention proxies:}
To test whether checking tracks public attention rather than internal
platform dynamics, we draw on two independent, non-Botometer sources.
The first is Google Trends search interest as weekly indices for September 2020 -- June 2023,
for the queries ``twitter bots'', ``botometer'', and ``elon musk twitter'', which serves as a Musk/Twitter-news confound check.

%Each term was pulled as a separate, single-term export (not downloaded jointly), so its 0--100 index is normalized against that term's own peak week and is not on a common scale with the other two; only within-series movement over time, not cross-series levels, is meaningful, which is why Figure~\ref{fig:events}b reports all three z-scored rather than on the raw 0--100 axis. Monthly figures aggregate the weekly index by an unweighted mean of the weeks falling in each calendar month, with no further renormalization. \textbf{[open: geography (worldwide vs.\ a specific country), search type (Web Search vs.\ a restricted category), and retrieval date need to be confirmed against the original query and added here]}.
The second is
a Pushshift~\cite{baumgartner2020pushshift} dump of Reddit's r/Twitter, a general Twitter-support subreddit, consisting of 191{,}289 submissions and
691{,}786 comments from April 2008 onward. We compute the weekly
share of non-bot-authored posts and comments containing
``bot''/``bots'' or ``fake'', after discarding its own
utility-bot traffic (any author whose name contains ``bot'').

%% file: sections/methods.tex
\section{Methods}

\begin{figure}[t]
\centering
\includegraphics[width=\columnwidth]{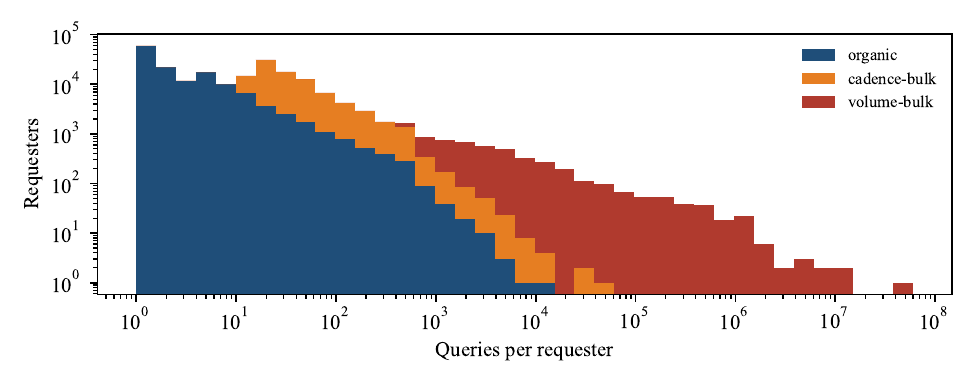}
\caption{Queries per requester (log--log), by filter class. }
\label{fig:requesters}
\end{figure}

\noindent\textbf{Separating organic from bulk traffic:}
Botometer serves two very different populations: researchers and
developers running programmatic sweeps over account lists, and
individuals checking specific accounts. Only the latter reflects public
suspicion. We remove programmatic traffic in two stages
(Figure~\ref{fig:requesters}). \emph{Volume:} a requester is bulk on
any day it issues more than 500 queries, and requesters that are ever bulk are
removed entirely, eliminating \nBulkIps{} of \nIps{} requesters but
\bulkShare\% of query volume. \emph{Cadence:} among the remaining requesters we additionally
remove those issuing more than 10 queries with a
\emph{median} gap between consecutive queries under 10 seconds -- a machine-like speed at which no human reads a result page. This
second stage removes \nCadenceIps{} requesters carrying \cadenceShare\% of the
remaining volume, confirming that low-volume scripted traffic is
substantial. The latter validates the former: \fdVolAlsoCadPct\% of volume-bulk requesters would also be caught by the
cadence criterion. We test the sensitivity of the paper's conclusions to these threshold choices in Appendix~A.

The final organic stream comprises \nOrganic{} queries
from \nOrganicIps{} requesters.
After filtering, we observe that \habitPctOneDay\% of organic
requesters appear on a single day only, and just \habitPctTenDays\%
are active on ten or more days. This usage profile is consistent with genuine dispersed human attention rather than a small number of persistent actors the volume/cadence stages failed to catch. Meanwhile, the \habitPctReturn\% who do return contribute \habitReturnCheckShare\% of all checks, indicating that a minority of requesters account for the majority of checks. This also indicates that most checks come from requesters checking accounts other than their own, ruling out the possibility that the majority of checks are simply users checking themselves.

\noindent\textbf{Unit of analysis:}
A single requester may re-query the same account many times, which
would inflate attention volume. Our unit is therefore the
\emph{check}: a unique requester--target--day triple, where target is
the checked account. A target's total number of checks is this
distinct-requester count summed across days: repeat checks by the same
requester count on later days, but not within the same day. This
deduplication removes \inflationShare\% of organic queries, yielding
\nChecks{} checks from an average of \requestersPerDay{} distinct
requesters per day across \nOrganicTargets{} organic targets, which
constitute \shOrganicTargets\% of the \nTargets{} accounts in the raw
query log.

\noindent\textbf{Joint Dataset:} We join the checked accounts in the Botometer query log with the accounts in the stream sample. By doing so, we collect additional account data such as the target's primary tweet language, account creation date, and activity record. \targetCoverage\% of organic targets (\targetMatched{}) appear in the
stream at least once.  We also collect daily mention counts of the accounts
involved in suspicion spikes, a measure of conversational visibility. %These are counted from every tweet in the stream sample (not the
%subsampled background below): all \nEvents{} suspicion-event
%accounts, at full 1\% resolution.
From this dataset we compute three covariates. The first is \emph{account age at check}, which is the duration between the check date and the target's profile creation date retrieved from the stream, available for \targetCoverage\% of targets. We also identify accounts with auto-generated profile handles (a name followed by a long digit string, e.g.,
\texttt{John95197683}), which correspond to the popular stereotype of a bot. Lastly, we compute \textit{modal tweet language} which is the plurality language the account uses in its tweets. Importantly, wherever we report ``bot score", we use the \textit{English score} if the account's modal tweet language is English and \textit{universal score} otherwise.

\noindent\textbf{Background Accounts:} We collect a sample of the stream as a baseline of the background tweeting population that will serve as a comparison group for the checked group. This background is itself a 1-in-100 subsample of the (already 1\%) stream for tractability, as capturing every author snapshot for the full stream's \archiveTweets{} tweets would be two orders of magnitude larger than needed. Even subsampled, it yields 19.4 million background accounts. %Target and spike coverage are not subsampled, only the background is; this ~100x density difference was noted in the now-cut bio/content-detail appendix where it affected comparisons.

As the latter is drawn from millions of accounts, every checked-vs-background comparison in the paper is
\textit{statistically significant} at $p<0.001$. We therefore
report effect sizes (ratios, percentage-point gaps) throughout rather
than repeating significance tests at each one, and let their
magnitude carry the comparison, unless otherwise stated.

%% file: sections/results.tex
% Assorted commented-out draft prose, cut findings, and cut table/figure
% blocks formerly inline in this file were moved to
% legacy/results_tex_removed_comments.txt during the 2026-09-16 cleanup
% pass (includes the Table~\ref{tab:descriptives} descriptive-statistics
% table and its restore instructions).

\section{RQ1: Which Accounts Attract Suspicion?}

We characterise checked accounts along three dimensions: how often
and when they are checked, their profile characteristics, handles, and what
their bios and tweets say and look like. We compare targets with the background population.

\noindent\textbf{How Often?} \pctOnce\% of
checked accounts are checked exactly once, the 100 most-checked
accounts absorb only \topHundredShare\% of all checks, the top
1{,}000 only \topThousandShare\%, and the top 10{,}000 only
\topTenThousandShare\%.
This suggests that the public checks are not centered on a handful of prominent accounts and attention to targets is extraordinarily dispersed.

\noindent\textbf{When?}
We test whether checking is timed to the target's posting activity. For each target, we compare activity on the day of the first check
against a placebo: the same measurement averaged on $K$ other days for the
same account, drawn at random from the archive's full coverage
(September 2020 -- January 2023) and excluding a $\pm$14-day buffer
around the actual check so the event and its immediate lead-up or
aftermath cannot contaminate the ``ordinary day'' baseline. We test for ``elevated'' activity which means the target's tweet volume that day was at least
twice its own $\pm$30-day baseline rate. For $K$ equals 24 (robustness to this choice is in Appendix~B), the target appears active in the
1\% sample on
\burstActivePct\% of check days, against \burstPlaceboActivePct\%
of placebo days, and is elevated on \burstObsElevatedPct\% of check
days versus \burstPlaceboElevatedPct\% of placebo days. Because a
1\% sample severely dilutes single-day activity, these gaps are lower
bounds; they corroborate that suspicion is typically triggered by
something the account just did. Elevated activity is not confined to the day of the check, it persists the day after the check
(\burstPostOneActivePct\% active, \burstPostOneElevatedPct\% elevated) and is still above
the placebo baseline a week later (\burstPostWeekActivePct\% active,
\burstPostWeekElevatedPct\% elevated, against
\burstPlaceboActivePct\%/\burstPlaceboElevatedPct\% for the placebo
date). Thus, checking tends to fall within a sustained active
period rather than reacting to a single spike that
quickly subsides.

\begin{figure}[t]
\centering
\includegraphics[width=\columnwidth]{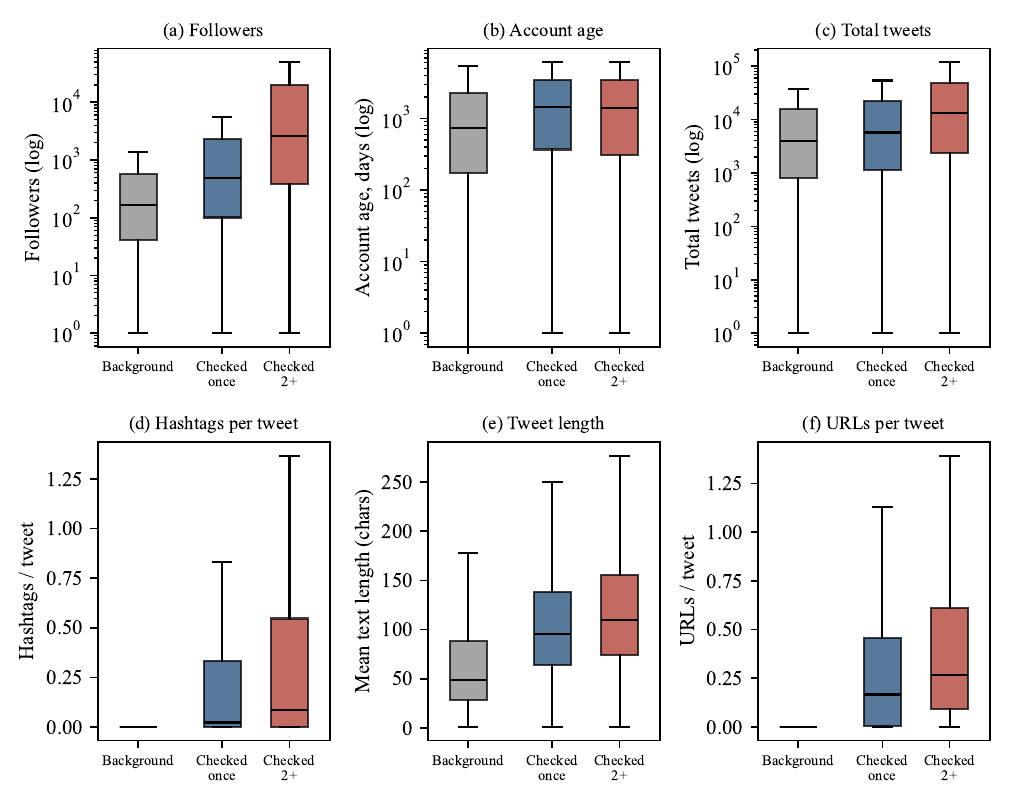}
\caption{Background population vs.\ checked accounts, split into
checked-once and checked-2+
The background box is a flat line at zero in (d) and (f): \tsBgHashtagZeroPct\%
of background accounts never use a hashtag and \tsBgUrlZeroPct\% never
post a URL, so both the 25th and 75th percentiles sit at zero.}
\label{fig:dispersion}
\end{figure}

\noindent\textbf{Profile Characteristics:} Checked accounts may have
disproportionately higher reach as they are older, have more followers and tweet more. Compared with the background population, checked accounts have four times the median follower count (591 vs.\
154), with an even larger disparity in mean follower counts (41,208 vs. 1,992). They are thirteen times as likely to be verified (4.6\% vs.\
0.35\%),
and are over-represented among higher-follower accounts
(Figure~\ref{fig:dispersion}a), yet mega-accounts ($>$1M followers)
remain only 0.66\% of targets. Suspicion is aimed at a wide,
largely non-elite set of accounts, but weighted towards the more
visible end of ordinary participants. This gap is not
uniform: repeat-checked targets (2+ checks)
have over five times the median followers of once-checked targets
(\multiMedFollowers{} vs.\ \onceMedFollowers{}).

The median target was \targetMedianAgeExact{} days old at first check, \emph{older} than the background tweeting population's \bgMedianAge{} days. The gap holds by creation date too, median \targetCreationMed{} for targets vs.\ \bgCreationMed{} for background, about 20 months earlier.
Only \targetPctYoungExact\% were 30 days old or younger, \emph{below}
the background rate of \bgPctYoung\% (Figure~\ref{fig:dispersion}b). The public is not primarily
suspicious of freshly created accounts; suspicion attaches to accounts
old enough to have accumulated a visible behavioural record.
Unlike followers, this age gap does not widen with repeat
checking: once- and repeat-checked accounts have almost identical
median ages (\onceMedAge{} vs.\ \multiMedAge{} days), both simply
older than background.

Checked accounts are also more prolific: a median
\targetMedianStatuses{} lifetime tweets against \bgMedianStatuses{}
for the background, a \statusesRatio$\times$ enrichment
(Figure~\ref{fig:dispersion}c). Repeat-checked targets post more than double the lifetime
tweets of once-checked targets (\multiMedStatuses{} vs.\
\onceMedStatuses{}). Thus, suspicion attaches to accounts old and active enough to have accumulated a visible track record.

\noindent\textbf{Default Handles:}
\pctDefaultHandle\% of checked accounts carry Twitter's auto-generated
handle pattern (name plus long digit suffix), the visual shorthand for
``bot'' in popular discourse. Among accounts active in the 1\% stream
over the same period, the background rate is \bgHandleUser\%
(\bgHandleTweet\% tweet-weighted): checked accounts (\pctDefaultHandle\%)
carry the pattern at essentially the same rate. This suggests that the public is more likely to check accounts when they encounter certain behaviour instead of stereotypical bot cues like generated handles.

\noindent\textbf{Languages:}
Checked accounts skew heavily
towards English: \langEnChecked\% of matched targets tweet
predominantly in English, against \langEnBg\% of the ambient stream.
Japanese, the stream's second language at \langJaBg\%, accounts for just \langJaChecked\% of checked targets. Bot suspicion, as instrumented through Botometer, is thus largely an anglophone
practice: the tool's interface and its public visibility are
English-language phenomena, and the accounts it is aimed at follow
suit.

\noindent\textbf{Content Markers:} Checked accounts are not blank
profiles: they are somewhat \emph{less} likely to have an empty bio
than the background (\ctBioEmpty\% vs.\ \cbBioEmpty\%,
\ratioBioEmpty$\times$). To characterise what checked accounts
present themselves as, we employ the vocabulary they use in their bios and tweets. The measurement has
three stages, applied identically to four content types: bio
words, bio hashtags, tweet words, and tweet hashtags. First, every token is scored for how sharply it
separates checked accounts from the background accounts, by weighted
log-odds with an informative Dirichlet prior
\citep{monroe2008fightin}, a symmetric \emph{distinctiveness}
ranking in which a token can be characteristic of either
population. As checked accounts are overwhelmingly English (\langEnChecked\% against \langEnBg\% of the background), two word-level rankings are computed on English-modal accounts only to prevent the free-text ranking from reporting the language gap instead of words that distinguish checked and background accounts.
As hashtags are self-delimiting topical
tokens, they are less likely to carry such a confound, so both hashtag rankings use the
full multilingual population.

Second, the 250 most checked-distinctive and 250 most
background-distinctive tokens of each content type are classified using Claude Sonnet 5. We first did an open-coding pass where the LLM was not fed any classes. We then merged similar classes to categories and kept the prevalent ones. Finally, we did another classification pass where tokens are either classified into one of the prevalent categories or into other.
Third, each category is measured as the union of its terms, over the
full population for the two hashtag content types and over English-modal
accounts only for the two word content types. For each category \(c\) with coded term set \(T_c\), let \(n_c^{\mathrm{chk}}\) and \(n_c^{\mathrm{bg}}\) denote the numbers of checked and background bios, respectively, that contain at least one term \(t \in T_c\). We define the enrichment ratio as \(R_c = (n_c^{\mathrm{chk}}/N^{\mathrm{chk}})/(n_c^{\mathrm{bg}}/N^{\mathrm{bg}})\).

The denominators \(N^{\mathrm{chk}}\) and \(N^{\mathrm{bg}}\) depend on the content type. For bio/tweet hashtags, they are the full account/tweet populations, and for bio/tweet words, they are the English modal account/tweet populations.

\begin{figure}[t]
\centering
\includegraphics[width=\columnwidth]{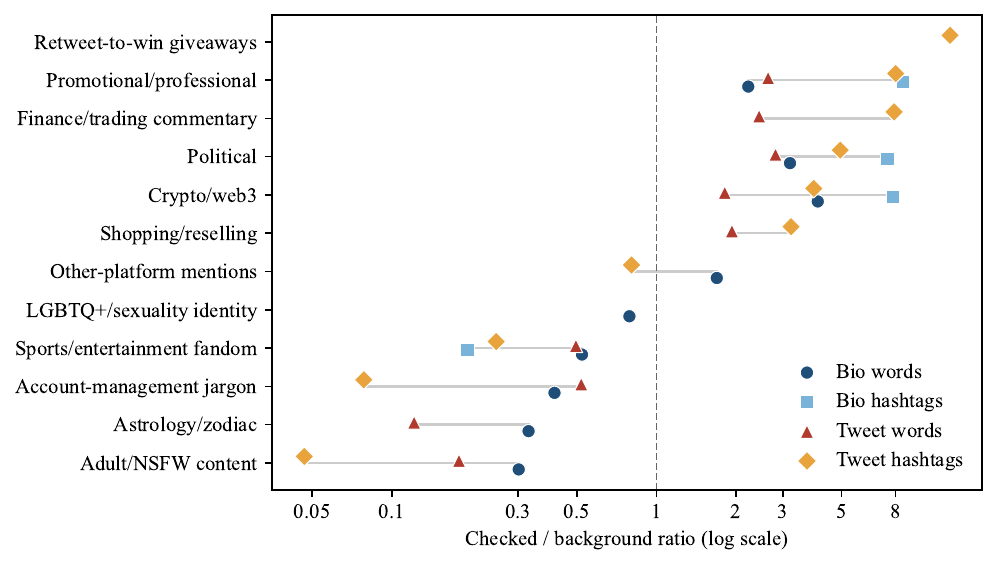}
\caption{Checked-vs-background enrichment ratio for
the same categories measured on four content types.}
\label{fig:biomarkers}
\end{figure}

Figure~\ref{fig:biomarkers} shows the categories and content types. We observe that
the same categories draw suspicion regardless of which of the four
content types is considered.
Political, crypto/web3, promotional/professional, finance/trading
commentary, shopping/reselling, and
retweet-to-win giveaways are enriched among checked accounts in every
content type where they are measured (1.8--12.9$\times$ background), and sports/entertainment
fandom, astrology, account-management jargon (e.g., \emph{dni} ``do
not interact,'' \emph{pfp}, \emph{acc}), and adult/NSFW content are
depleted in every content type where they are measured
(0.05--0.52$\times$). LGBTQ+/sexuality identity is roughly at parity
(0.79$\times$, its only measured level). The one exception among the
enriched/depleted categories is other-platform mentions, which flip sign: checked-enriched in bio words (1.7$\times$) but
background-enriched in tweet hashtags (0.8$\times$). Upon manual inspection, we observe that these are two different use cases: mentioning other platforms in a bio points to self-promotion, spam behaviour, e.g., ``Join my Telegram channel'' while tweet hashtags appear to be routine auto-cross-posting, e.g., when a user shares an Instagram post on Twitter, the hashtag \#instagram is inserted into the post.

\begin{figure}[t]
\centering
\includegraphics[trim=0 0 0 13pt, clip, width=\columnwidth]{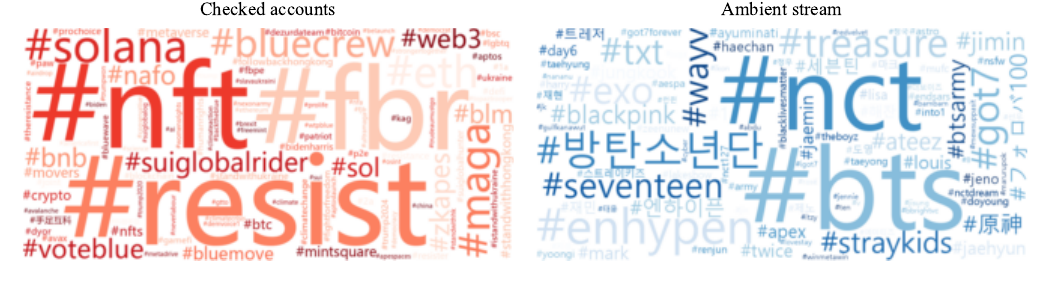}
\caption{Checked (left) vs. ambient (right) bio hashtags, sized by weighted log-odds $|z|$.}
\label{fig:biohashtagcloud}
\end{figure}

Figure~\ref{fig:biohashtagcloud} makes the split visible at the bio-hashtag level: \#nft, \#resist, \#fbr, and \#maga on one side against \#bts, \#nct, \#enhypen, and the rest of the K-pop cluster (in both Latin and Korean script) on the other. This asymmetry is consistent with public suspicion tracking what an account \emph{advocates} (promotion, partisanship, crypto) rather than how automated its behaviour looks, although the ambient-stream comparison cannot tell whether these topics increase suspicion once encountered or are simply encountered more often by Botometer users. This contrast is particularly notable because fandom accounts are the platform's most numerous high-volume, repetitive, pseudonymous genre, the population that may most resemble the folk image of a bot, and they are among the least checked. This is consistent with prior work on ``bot'' as a move in political argument rather than a technical diagnosis \citep{assenmacher2023bot}. An alternative reading is that fandom accounts are checked less because the public already assumes they are automated and sees no need to verify. Usage of default handles, the most literal folk bot cue, argues against this: checked accounts carry the auto-generated pattern at essentially the background rate, so settled belief driven by an obvious stereotype cue does not generally suppress checking. A third reading is a filter bubble rather than a suspicion gap: fandom accounts, K-pop especially, target a global audience, and checking is itself an anglophone practice, so an anglophone checker may simply rarely encounter fandom content to check in the first place.

\noindent\textbf{Tweet Structure:}
Checked accounts' original
tweets are visibly more promotional in construction than the
background's. They are more likely to carry a URL (\ctUrlShare\% vs.\
\cbUrlShare\%), twice as likely to carry a hashtag (\ctTagShare\%
vs.\ \cbTagShare\%), more mention-heavy (\ctMentionShare\% vs.\
\cbMentionShare\%), and somewhat less retweet-heavy (\ctRtShare\%
vs.\ \cbRtShare\%). The full per-account distributions
(Figure~\ref{fig:dispersion}d--f) show this is not a step change but
a graded one, following the same background/checked-once/checked-2+
split as the profile covariates in that figure's first three panels:
median hashtags per tweet rises from \tsHtBgMed{} (background) to
\tsHtOnceMed{} (checked once) to \tsHtMultiMed{} (checked 2+), median
tweet length from \tsLenBgMed{} to \tsLenOnceMed{} to
\tsLenMultiMed{} characters, and median URLs per tweet from
\tsUrlBgMed{} to \tsUrlOnceMed{} to \tsUrlMultiMed{}.

\section{RQ2: When Does Suspicion Become Collective}

RQ1 characterised suspicion as dispersed: most targets are checked once, by one requester, and never again. Here we investigate the contrasting case in which suspicion concentrates sharply on particular accounts. We first define and quantify these collective suspicion spikes and relate their timing to overall checking volume and to attention outside the platform. We then turn to the accounts themselves, examining their conversational visibility and what they were doing when suspicion converged on them.

\noindent\textbf{Collective suspicion spikes:}
We define a \textit{suspicion spike} as an account receiving checks from at least $k$ distinct requesters on a single day. Such spikes capture collective, temporally concentrated attention that is typically associated with the account's visibility in an ongoing conversation.
We set $k=5$: high enough that a handful of unrelated people
happening to check the same moderately-visible account on the same
day is an implausible explanation,
while low enough to retain obscure, chronically-flagged accounts
alongside celebrities (Table~\ref{tab:events}). We test this choice by replicating the paper's conclusions at $k=3$ through $k=12$ (Appendix~C), showing that the findings hold throughout.

\begin{figure*}[t]
\centering
\includegraphics[width=\textwidth]{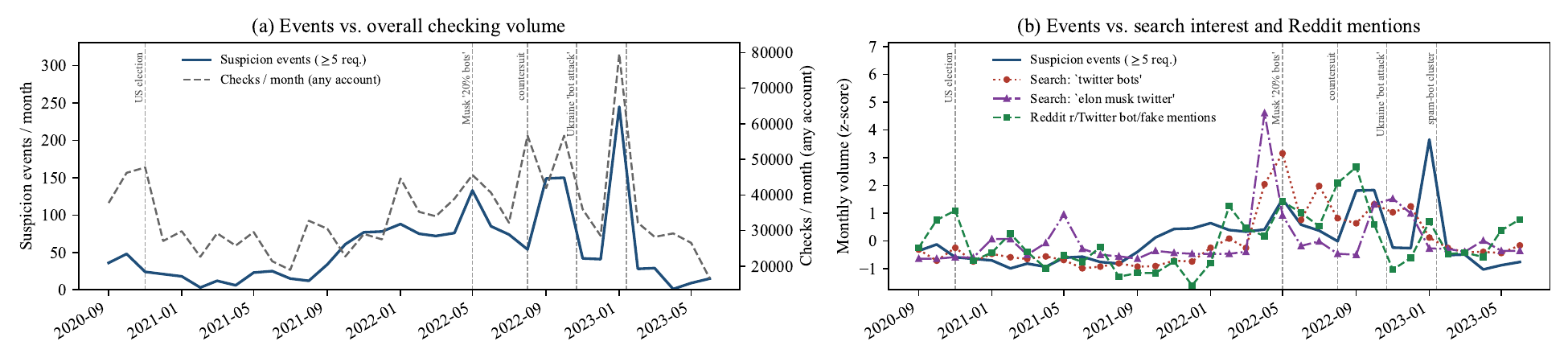}
\caption{(a) Monthly suspicion spikes (left axis) against overall
checking volume (right axis): the two follow
each other loosely ($r=\evChecksEventsR$). Dashed lines mark Elon Musk-related events, the October 22--23, 2022
Ukraine-conflict-adjacent cluster, and the January 12--15, 2023
spam-account cluster discussed in the text. (b) Monthly suspicion spikes against Google search
interest in ``twitter bots'' and ``elon musk twitter'', and against
the r/Twitter bot/fake mention share (all z-scored as raw search interest is a 0--100 index). Spikes correlate more closely with bot-specific
search interest ($r=\mediaEventsTwitterBotsR$) than with general
Musk/Twitter search interest ($r=\mediaEventsMuskR$), visible in May 2022, where ``elon musk twitter'' searches spike far above
everything else while spikes rise only modestly. The Reddit mention
share correlates more weakly, and not significantly, ($r=\mediaEventsRedditR$,
$\rho=\mediaEventsRedditRho$).}
\label{fig:events}
\end{figure*}

We observe \nEvents{}
suspicion spikes. The spike series
(Figure~\ref{fig:events}) tracks platform history closely: monthly
spike counts run from 3 to 34 through most of 2021, climb from October
2021 (61), stay elevated through
2022 (mostly 41--150 per month), and vanish with the API closure. The single largest collective spike of the entire record is
\emph{Elon Musk himself}: his account drew 217 distinct
requesters on August 5, 2022 and 161 the next day, the week his
countersuit over Twitter's bot-prevalence claims dominated the news.
An earlier surge of comparable scale followed his ``20\% bots''
claims in May 2022.

\shEventAccounts{} accounts ever drew a spike: \evSpikeOncePct{}\% just once, \evSpikeMidPct{}\% two to five times, and \evSpikeTailPct{}\% more than five, with Musk's account (\evSpikeMaxCount{}) an extreme outlier. Table~\ref{tab:events} shows the top accounts, ranked by the size of their single largest spike. Ordinary accounts are described by category for privacy reasons. They are heterogeneous: partisan accounts in three languages, a
Spanish-language media channel, an academic (outside of the Botometer team), a high-volume Japanese
account, and a personal account with a few dozen followers.

\noindent\textbf{Vs. Overall Checks:} Monthly spike counts correlate
with overall monthly checking volume (every organic check on any account,
regardless of whether it is part of a spike) ($r=\evChecksEventsR$,
$\rho=\evChecksEventsRho$, $n=\evChecksEventsN$ months) but far from
proportionally (Figure~\ref{fig:events}). Spikes are essentially flat
through the record's first thirteen months (\evPreBreakEventsMean{}
per month on average, September 2020--September 2021) and then step
up roughly \evPrePostEventsGrowth\hspace{0pt}--fold for the rest of the
record (\evPostBreakEventsMean{} per month, October 2021 onward),
while overall checking volume shows no comparable break
(\evPreBreakChecksMean{} vs.\ \evPostBreakChecksMean{} checks/month).
Even the single record month (January 2023, the peak for both series)
diverges: checks rose \evPeakChecksGrowth$\times$ over the prior month,
spikes \evPeakEventsGrowth$\times$. Targeted collective suspicion
therefore shows a distinct temporal shift not mirrored by regular
checking.

Spikes diverge from overall checks through several distinct
episodes. Spikes rise several-fold in October 2021 and remain elevated for the following year. A cluster on October 22--23, 2022 concentrates
on Ukrainian- and Russian-language political accounts, including
presidential adviser Mykhailo Podolyak.
This coincides with a renewed public dispute between Musk and Ukrainian officials as his Twitter acquisition neared completion. On October~3, Musk polled his followers on a proposed ``Ukraine--Russia Peace'' settlement~\cite{cnn2022muskukraine}. The proposal drew widespread opposition, including from Podolyak, who publicly countered with his own plan calling for Ukraine to recover its occupied territories \cite{dudik2022muskukraine}. As Musk's poll turned decisively against his proposal, he attributed the result to a ``bot attack'' \citep{cnn2022muskukraine}, invoking the same framing that had directed bot suspicion towards accounts in this community.

A larger cluster on January
12--15, 2023 has a different signature: 212 distinct targets, each
checked by only 5--15 requesters, overwhelmingly handles of the
auto-generated \texttt{name+digits} form associated with spam and
crypto-promotion bots. This is a decentralized, simultaneous bot-hunting
across many near-identical low-profile accounts rather than one
high-profile target drawing a crowd. Both clusters are dashed in
Figure~\ref{fig:events}a; the January one alone accounts for 220 of
that month's 245 spikes.

\noindent\textbf{Vs. Media Attention:}
We test whether checking activity is associated with independent measures
of public attention to bots. Monthly check volume correlates with Google
search interest in ``twitter bots'' at $r=\mediaChecksTwitterBotsR$ ($t=\mediaChecksTwitterBotsT$, Spearman
$\rho=\mediaChecksTwitterBotsRho$), with searches for ``botometer'' at $r=\mediaChecksBotometerR$ ($t=\mediaChecksBotometerT$,
Spearman $\rho=\mediaChecksBotometerRho$), and with the share of r/Twitter posts and comments
mentioning ``bot''/``fake'' at $r=\mediaChecksRedditR$ ($t=\mediaChecksRedditT$, Spearman $\rho=\mediaChecksRedditRho$)
(Figure~\ref{fig:events}). Monthly suspicion spikes correlate similarly
with the first two proxies, at $r=\mediaEventsTwitterBotsR$ ($t=\mediaEventsTwitterBotsT$, Spearman $\rho=\mediaEventsTwitterBotsRho$)
for ``twitter bots'' and $r=\mediaEventsBotometerR$ ($t=\mediaEventsBotometerT$, Spearman $\rho=\mediaEventsBotometerRho$) for
``botometer''; the Reddit correlation is weaker
($r=\mediaEventsRedditR$, $t=\mediaEventsRedditT$, Spearman $\rho=\mediaEventsRedditRho$).
General Musk/Twitter attention does not show the same pattern: search
interest in ``elon musk twitter'' correlates only weakly with suspicion
spikes ($r=\mediaEventsMuskR$, $t=\mediaEventsMuskT$, $\rho=\mediaEventsMuskRho$, $n=34$) and monthly check
volume ($r=\mediaChecksMuskR$, $t=\mediaChecksMuskT$, $\rho=\mediaChecksMuskRho$).

These monthly correlations show that bot-related attention and
Botometer activity tend to be higher during the same periods, but
they do not tell us whether the two rise and fall together from one month
to the next. To test this, we first-difference each series, replacing
each month's value with its change from the previous month, and correlate
these changes. The Google Trends associations then largely disappear:
``twitter bots'' falls to $r=\mediaEventsTwitterBotsDiffR$ for spikes
and $r=\mediaChecksTwitterBotsDiffR$ for checks, while ``botometer''
falls to $r=\mediaEventsBotometerDiffR$ and
$r=\mediaChecksBotometerDiffR$, respectively. Thus, the original monthly
correlations primarily reflect bot-related attention and Botometer
activity being higher during the same periods, rather than their
month-to-month changes moving together. These correlations are
nevertheless robust to excluding individual months, so they are not
driven by a single exceptional month. The Reddit association is less
robust: it is sensitive to how weekly data are aggregated to months and
disappears at daily resolution ($r=\dailyChecksRedditR$; changes from
one day to the next, $r=\dailyDiffRedditR$). Overall, external
bot-related attention and Botometer activity tend to be higher during
the same periods, but we find little evidence that their short-run
fluctuations move together.

\begin{table}[t]
\centering
\small
\caption{Top ten accounts with the largest collective suspicion spikes. Chk.: distinct
requesters on that account's single largest spike day (the ranking
criterion); Ev.: that account's total suspicion spikes.
Score: bot score at first check; Ment.: spike-day mention rate
relative to the account's own $\pm$30-day baseline in the stream
(``--'': the account is never mentioned in the sample,
``$>$'': mentioned only on the spike day)}
\label{tab:events}
\input{tables/events_case}
\end{table}

\noindent\textbf{Vs. Account Visibility:}
We find that collective suspicion is associated with elevated visibility for a sizeable minority of targets.
We track every spike account's daily mention count across the full
archive window, then for each spike compute a local baseline: the
median mention rate over the surrounding $\pm$30 days, excluding the
2 days immediately around the spike itself so the elevation cannot
bias its own baseline. Many targets are essentially never mentioned
outside their own spike days, so this baseline is exactly 0 for a
large share of them, which would leave a raw mention-rate/baseline
ratio undefined; we add 1 to both the spike-day mention count and the
baseline before dividing, and aggregate
account-wise: for each of the \evAccountsN{} accounts we pool its own
mention and baseline totals across all its spikes, smooth once, then
average across accounts equally.

The mean elevation is \evElevMean$\times$ baseline but it is carried by a minority. \evElevLeOnePct\% of accounts show no elevation at all
(ratio $\leq$1$\times$), \evElevMidPct\% show some but stay below a
doubling (1--2$\times$), and \evElevGeTwoPct\% at least double their
baseline mention rate. The mean elevation drops from \evElevMean$\times$ on
the spike day to \evElevPostOneMean$\times$ the day after and
\evElevPostWeekMean$\times$ a week later.

\noindent\textbf{What spike targets were doing:}
The targets' own sampled tweets show what collective suspicion reacts
to (Table~\ref{tab:archetypes}). 43.9\% are active according to the stream sample
itself and we classify what they were doing from that content below. A further
\evbActiveUncapturedPct\% are active only according to Botometer's
own check-time record (most recent tweet and tweets per day): the target had
tweeted within days of the spike at a substantial rate, just not
visibly enough for a sample that only captures 1\% of all tweets to
catch. The remaining 14.3\% are not confirmed active by either
source: \evbDormantPct\% look genuinely
dormant (no tweet in the last 30 days, or averaging fewer than
one tweet every 10 days according to the Botometer log), and
\evbLowActivityPct\% fall between the active and dormant
bounds and stay ambiguous.
We classify the accounts from the first group according to their behaviour in a multilabel fashion: mostly retweets, mostly replies, hashtag-stuffing, activity burst and others. Table~\ref{tab:archetypes} shows the classification rules and the distributions. Collective suspicion falls on amplifiers, consistent with the retweet-bot markets characterised in prior work~\cite{elmas2022retweetbots}: \evbMostlyRetweetsPct\% of targets were predominantly
\emph{retweeting} (retweet share $>0.5$). Mostly-replies (reply share $>0.5$) spikes are usually the targets arguing in long threads at elevated rates and draw the \emph{lowest}
bot scores of any archetype (median 0.48). This may suggest that crowds that check a combative
interlocutor are usually checking a human. Hashtag-stuffing
($\geq$2 hashtags/tweet) accounts for just \evbSpamPct\% of spikes. This may suggest that
the accounts that trigger collective suspicion rarely look like
naive spammers. A further \evbActiveElevatedPct\% post at least
twice their own baseline rate that day, but this mostly overlaps
with the categories above rather than marking a distinct behaviour on
its own: only \evbBurstOnlyPct\% show elevated activity with none of
the other three signatures.

\begin{table}[t]
\centering
\small
\caption{Behaviour of spike targets on their spike days. Each spike is assigned the most specific
category its evidence supports: where the target's tweets appear in the
1\% sample we classify the content directly, and where they do not the
account's activity level at check time determines the category
(definitions in column two). Med.\ score
is the target's bot score at first check.}
\label{tab:archetypes}
\resizebox{\columnwidth}{!}{\input{tables/event_archetypes}}
\end{table}

\section{RQ3: What Comes of Suspicion?}

\noindent\textbf{Account Disappearance:}
We hypothesise that if public suspicion carries signal, checked accounts
would be more likely to disappear sooner. Accounts can disappear through
platform suspension or dormancy due to undisclosed platform mechanics
such as phone verification~\cite{elmas2023suspension}.
The stream join permits testing both without distinguishing them.
For each matched target active in the stream during the 90 days before
its first check, we ask whether it ever appears again between 14 and 104
days after the check; the two-week grace period discards the activity
burst that typically triggers checking. Controls are targets whose first
check lies at least 120 days in the future: accounts that will eventually
attract suspicion but have not yet been checked. We use this control
because background accounts' activity does not match that of checked
accounts, as shown in RQ1. Future-checked accounts are more comparable
because they eventually attract scrutiny but have not yet
been checked at observation.

Across monthly cohorts
(\disTreatedN{} eligible checked accounts), \disTreatedPct\% of
checked accounts disappear from the stream, against \disControlPct\%
of not-yet-checked controls (ratio \disRatio{}, $z=\disZ$). The
difference is not immediate: at 30 days checked accounts are
\emph{less} likely to have vanished than controls (28.8\% vs.\ 30.5\%,
ratio 0.94, since a checked account was by construction active enough
to attract suspicion and typically keeps tweeting right after). The
gap opens only at longer horizons: 20.1\% vs.\ 19.2\% (ratio 1.05) at
60 days, 16.5\% vs.\ 14.2\% (ratio 1.16) at 90 days, and 14.5\% vs.\
11.2\% (ratio 1.29) at 120 days. This may reveal a lag in platform enforcement: the accounts only get suspended or cease activity over a long follow-up period.
As checked accounts are more active before observation, we stratify the comparison by pre-window activity level. Checked accounts disappear more often at every activity level. When we give the control group the same activity-level mix as the checked accounts, its disappearance rate actually falls to \disControlAdjPct\% (from a raw 14.3\%). The treated-to-control ratio widens accordingly, from \disRatio{} to 1.39. (Bucket-by-bucket detail and the standardization formula in Appendix~D).

\noindent\textbf{Suspension:} To investigate whether stream disappearance
reflects platform enforcement, we queried the current (August 2026) status
of \usSampleN{} sampled targets directly on the platform: \usGoneN{} that
had disappeared from the stream after their first check and
\usPresentN{} that remained present. Stream disappearance is not
suspension: accounts that remained present are suspended somewhat
\emph{more} often than disappeared ones (\usPresentSuspPct\% vs.\
\usGoneSuspPct\%, $p=\usSuspP$).

\noindent\textbf{Dormancy:} Accounts can go quiet without ever being formally suspended, so we also check their activity directly. Of the
\goneAliveActN{} stream-disappeared accounts confirmed alive above,
\goneAliveUnavailPct\% had already become deactivated and a further \goneAliveActNoVisiblePct\%
returned no visible tweets at all (protected, empty, or too sparse to
surface). Of the \goneAliveActWithTweetsN{} with a visible timeline, \goneAliveDormantYearOnePct\% cease all activity within a year of disappearing from the stream and never tweet again,
\goneAliveDormantYearTwoPct\% within two years, and
\goneAliveDormantYearThreePct\% within three.
By contrast, accounts that never disappeared from the stream go dormant at
roughly a third the rate over the same check-relative measure:
\presentDormantYearOnePct\% within a year, \presentDormantYearTwoPct\%
within two, \presentDormantYearThreePct\% within three.

\noindent\textbf{Vs. Suspicion Targets:}
We queried the current status of all 801 collective suspicion targets.
41.2\% are unreachable, either through suspension or deletion/deactivation.
Their suspension rate is \evStatusSuspPct\%, compared with \baseSuspPct\% in the general
sample of 5,000 accounts described earlier ($z=\evStatusSuspZ$, $p\evStatusSuspP$).
As that general sample deliberately contains equal numbers of
stream-disappeared and stream-present accounts, we reweight it to the
population's actual \disTreatedPct\% stream-disappearance rate. The resulting
estimated suspension rate is \baseSuspReweightedPct\%, still substantially below the
\evStatusSuspPct\% observed among collective-suspicion targets
($z=\evStatusSuspReweightedZ$, $p\evStatusSuspReweightedP$). This difference is also robust to adjustment for observed account
characteristics. In a weighted logistic regression controlling for
Botometer score, log-transformed follower and status counts, a
bio-based political-content flag, and calendar period, collective-suspicion
targets have \susAdjOR{} times the odds of suspension relative to the general
sample (95\% CI \susAdjORLow{}--\susAdjORHigh{}, $z=\susAdjZ$, $p\susAdjP$).
Adjustment changes the estimate very little: the corresponding
unadjusted odds ratio in the same complete-case sample is \susRawOR{}
(95\% CI \susRawORLow{}--\susRawORHigh{}).
Deletion/deactivation, by contrast, is not elevated: \evStatusGonePct\% of suspicion-spike accounts are gone this way, compared with \baseGonePct\% of the general sample ($z=\evStatusGoneZ$, $p=\evStatusGoneP$).
Their dormancy profiles are similar. Among collective suspicion targets that remain reachable, \eventDormantYearOnePct\% never tweeted again within a year of their most recent spike, compared with \goneAliveDormantYearOnePct\% of the \goneAliveActWithTweetsN{} stream-disappeared accounts with a visible timeline and \presentDormantYearOnePct\% of the \presentActWithTweetsN{} non-stream-disappeared ones with a visible timeline.

Public suspicion is thus associated with subsequent withdrawal from
the conversation: checked accounts are more likely to go quiet after
being checked. When suspicion is collective, the association extends to
suspension as well, with collectively suspected accounts showing a higher
subsequent suspension rate. These associations do not establish that
suspicion itself causes either outcome.

\begin{figure}[t]
\centering
\includegraphics[width=0.9\columnwidth]{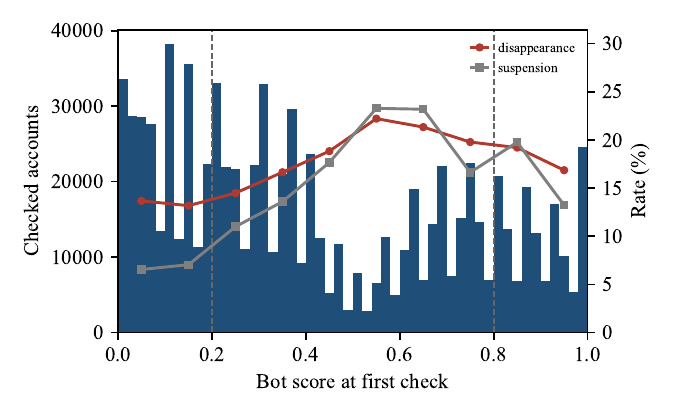}
\caption{Bot scores at first check. Lines show the stream-disappearance and
suspension rates by score bin. Dashed lines mark the conventional
human-like ($\leq 0.2$) and bot-like ($\geq 0.8$) thresholds.}
\label{fig:scorehist}
\end{figure}

\noindent\textbf{Scores:} We report the score returned at a
target's \emph{first} organic check throughout unless otherwise stated. Targets checked more than once (17.6\%) could
alternatively be summarised by the mean of all their organic-check
scores, and the two conventions usually agree closely. Across
\scoreN{} checked accounts, the median first-check score is
\scoreMedian{}; \scoreLowPct\% of targets score in the
human-like range ($\leq 0.2$) and only \scoreHighPct\% score
bot-like ($\geq 0.8$) as shown in Figure~\ref{fig:scorehist}. Collective suspicion is more strongly associated with higher Botometer scores
than individual suspicion: accounts that ever drew a suspicion spike
have a median first-check score of \scoreEventMedian{} and mean-check score of 0.56.
\scoreEventHighPct\% score
bot-like ($\geq 0.8$), against \scoreHighPct\% of checked accounts
overall. The scores also cohere
with subsequent stream
disappearance: disappearance within three
months of the first check rises monotonically across score quartiles,
from \scoreGoneQOne\% in the lowest (scores $\leq \scoreQOneMax$) to
\scoreGoneQFour\% in the highest ($\geq \scoreQFourMin$).
Figure~\ref{fig:scorehist}'s red line plots this disappearance rate
by score bin directly (\scoreDisRateMin--\scoreDisRateMax\%, rising
then leveling off past the midpoint). The grey line shows
same-bin suspension rate from the \usSampleN{}-account sample, reweighted by the true share of stream-disappeared
vs.\ not-disappeared accounts in each bin. The
reweighted suspension rate (\scoreSuspRateMin--\scoreSuspRateMax\%)
follows the disappearance rate closely.

\begin{figure}[t]
\centering
\includegraphics[width=0.9\columnwidth]{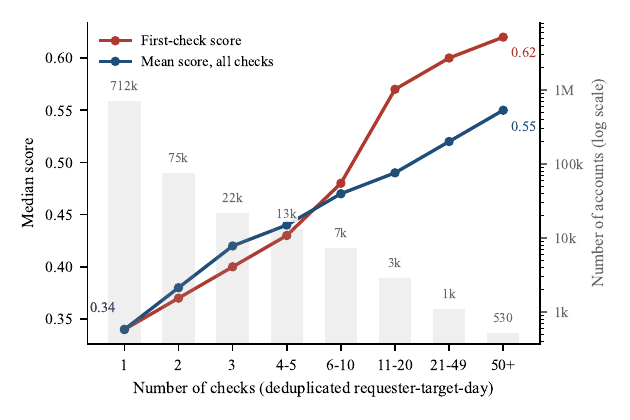}
\caption{Median first-check \& mean score by number of checks (left-axis), sample size (bars, right-axis).}
\label{fig:ncheck}
\end{figure}

\noindent\textbf{Scores vs. \#Checks:} We also observe that repeatedly-checked accounts score higher.
Using deduplicated checks to group the \ncN{} organic targets, we observe that median
first-check score rises monotonically from \ncOnceMed{}
(once-checked) to \ncTopMed{} for the \ncTopN{} targets checked
\ncTopLabel{} times (Figure~\ref{fig:ncheck}; $\rho=\ncRho$; averaging
over all checks gives the same order, $\rho=\ncMeanRho$). This suggests that repeat checking over time
may be a tractable chronic-suspicion proxy.

\begin{figure}[t]
\centering
\includegraphics[width=\columnwidth]{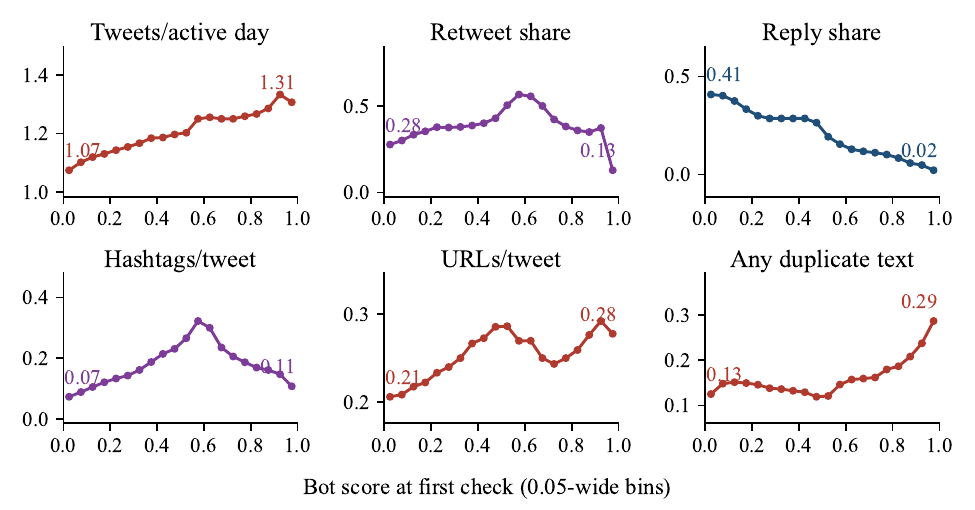}
\caption{Median stream behaviour of checked accounts by 0.05-wide bin
of the bot score at first check. Color marks each panel's overall shape:
rising end-to-end (red), falling end-to-end (blue), or peaking/dipping
in the mid-range above or below both endpoints (purple).}
\label{fig:behscore}
\end{figure}

\noindent\textbf{Scores vs. Behaviour:} We analyse \behScoredN{} matched
targets who have $\geq$5 sampled posts (including all types of
tweeting: original tweets, replies, and retweets), grouped into
0.05-wide score bins, by their posting behaviour and scores. As
Figure~\ref{fig:behscore} shows, posts (all types of tweets) per
active day rise steadily (1.07 to 1.31) and reply share falls
steadily (0.41 to 0.02). This may suggest that low-scoring targets argue, and do so at a
slightly lower posting rate. Retweet share and hashtag density,
however, are not a straight-line intensification of ``more bot-like
behaviour'': both rise from the low end to a peak around score 0.6
(retweet share 0.28 to 0.57; hashtags per tweet 0.07 to 0.32), then
recede, and crash in the highest bin (score $\geq$0.95: retweet share
0.13, hashtags 0.11). The share of accounts authoring at least one duplicate text also gradually increases, from ~12\% among the
lowest-scoring accounts to nearly 29\% in the highest-scoring bin.

\noindent\textbf{User Feedback:} We employed the feedback log dataset and classified the reasons cited by users into a 31-category, 7-theme taxonomy employing Claude Sonnet 5: categories were open-coded per label, unified into
one canonical taxonomy across labels, then every text reclassified
against it. We validated a subset of 120 data points (40 per label) against two independent
models: Cohen's $\kappa$ (granular/theme level) is 0.80/0.86 with Opus 5 and 0.77/0.80 with GPT-5.6, substantial to
almost-perfect chance-corrected agreement by
\citet{landis1977kappa}'s benchmarks. Table~\ref{tab:fbtheme} reports the
resulting category shares by label.
\emph{Human} feedback is overwhelmingly
first-person identity assertion: self-identity, ownership, and
personal-connection claims alone account for the large majority of
human free text, versus a small minority for bot or cyborg. \emph{Bot}
and \emph{cyborg} feedback is the opposite, evidence-dense third-person
argument citing posting rate, political content, and network
coordination.

Several categories name evidence outside the feature set of traditional per-account classifiers, e.g., network coordination (8.6\% of all feedback, 26.6\% of cyborg feedback),
which requires comparing multiple accounts against one another, the same kind of cross-account signal that identifies mass follow-back reciprocity schemes~\citep{elmas2024followback}. Account takeover/hijack (1.4\%, 7.0\% of cyborg) implies a behavioural
discontinuity within one account's own history that a single
current-snapshot score cannot represent, the same discontinuity a hijacked and repurposed account would show~\citep{elmas2023repurposing}; recent work modelling behavioural \emph{change} over an account's lifetime exists precisely
because static classifiers miss it \citep{ariyarathne2026behavior}.
DM (direct message) script pattern (0.4\%, 1.5\% of bot judgements), naming
automation observed in private messages, is invisible to any
classifier built on public API data. The public's
free-text justifications thus supply evidence a bot
detector may have missed.

\begin{table}[t]
\centering
\small
\caption{Reason categories by theme \& bot label (multilabel, percentages need not sum to 100 within a
column). ``All'' pools all labels' texts before computing the
percentage.}
\label{tab:fbtheme}
\resizebox{\columnwidth}{!}{%
\setlength{\tabcolsep}{3pt}%
\input{tables/feedback_text_theme_table}}
\end{table}

%% file: tables/events_case.tex
% Auto-generated by analysis/events_table.py -- do not edit by hand.
% Selection: largest event per account, top 10 accounts by checkers.
\footnotesize
\resizebox{\columnwidth}{!}{%
\begin{tabular}{@{}lcrrrrl@{}}
\toprule
Acct. & Lg. & Chk. & Ev. & Score & Ment. & Event-day Behaviour \\
\midrule
@elonmusk & en & 217 & 133 & 0.30 & 0.9$\times$ & No sampled tweets \\
Partisan & en & 70 & 2 & 0.87 & 9.5$\times$ & Mostly retweets \\
Academic & en & 36 & 2 & 0.11 & -- & No sampled tweets \\
Anti-fascist & es & 26 & 2 & 0.15 & $>$ & Mostly replies, Activity burst \\
Political & -- & 25 & 1 & 0.91 & 1.0$\times$ & No sampled tweets \\
Ukrainian & en & 22 & 1 & 0.20 & -- & No sampled tweets \\
Political & fa & 20 & 5 & 0.73 & 6.8$\times$ & Mostly replies, Activity burst \\
Media & es & 20 & 1 & 0.31 & 8.3$\times$ & No sampled tweets \\
Auto-gen. & es & 19 & 31 & 0.70 & 3.7$\times$ & Mostly retweets \\
High-vol. & ja & 19 & 1 & 0.20 & -- & Mostly replies \\
\bottomrule
\end{tabular}%
}

%% file: tables/event_archetypes.tex
% Auto-generated by analysis/rq1_event_content.py -- do not edit by hand.
% Rows above the rule are classified from sampled tweet content and are
% multilabel (an event-day may carry more than one signature, so counts
% do not sum to the 803-event active pool); rows below the rule are
% mutually exclusive and come from Botometer's check-time activity record.
\begin{tabular}{lp{3.6cm}rr}
\toprule
Event-day behaviour & Definition & Events & Med.\ score \\
\midrule
Mostly retweets & Retweet share $>0.5$ & 556 & 0.67 \\
Mostly replies & Reply share $>0.5$ & 91 & 0.48 \\
Hashtag-stuffing & $\geq$2 hashtags/tweet & 29 & 0.64 \\
Activity burst & Daily tweets $\geq 2\times$ the account's own baseline & 314 & 0.64 \\
No signature above & Active in the sample; none of the above met & 92 & 0.69 \\
\midrule
Active, sample-missed & No sampled tweets; last tweet $<$3 days before the event and $\geq$1 tweet/day & 765 & 0.57 \\
Low-activity / ambiguous & No sampled tweets; neither of the above bounds met & 158 & 0.69 \\
Dormant (genuinely quiet) & No sampled tweets; last tweet $\geq$30 days stale, or $<$0.1 tweets/day & 103 & 0.76 \\
\bottomrule
\end{tabular}

%% file: tables/feedback_text_theme_table.tex
% Auto-generated by analysis/feedback_text_categories_llm.py -- do not edit by hand.
\begin{tabular}{lrrrr}
\toprule
Category & All\% & Bot\% & Cyborg\% & Human\% \\
\midrule
\textbf{Posting behaviour} & \textbf{22.4} & \textbf{30.6} & \textbf{60.1} & \textbf{9.8} \\
\quad Posting rate / volume & 9.4 & 15.3 & 22.3 & 4.0 \\
\quad Content repetition (retweets, duplicates) & 5.9 & 12.0 & 13.5 & 1.7 \\
\quad Behavioural timing pattern & 4.5 & 6.4 & 17.7 & 0.5 \\
\quad Timezone/location mismatch & 2.5 & 1.2 & 13.5 & 0.1 \\
\quad Account age / inactivity anomaly & 4.8 & 1.2 & 12.4 & 4.2 \\
\textbf{Content} & \textbf{15.7} & \textbf{22.6} & \textbf{43.8} & \textbf{6.0} \\
\quad Political content & 10.7 & 12.9 & 28.7 & 5.4 \\
\quad Misinformation content & 1.3 & 3.1 & 3.0 & 0.1 \\
\quad Spam / scam / adult content & 4.5 & 6.0 & 18.6 & 0.4 \\
\quad Harassment / doxxing & 1.1 & 2.5 & 3.0 & 0.1 \\
\textbf{Account / network} & \textbf{14.7} & \textbf{22.3} & \textbf{39.5} & \textbf{5.7} \\
\quad Network coordination & 8.6 & 11.1 & 26.6 & 3.1 \\
\quad Follower/engagement mismatch & 2.6 & 5.1 & 5.6 & 1.0 \\
\quad Profile identity anomaly & 5.8 & 8.2 & 18.1 & 1.7 \\
\textbf{Interaction} & \textbf{2.6} & \textbf{2.3} & \textbf{1.4} & \textbf{3.1} \\
\quad DM script pattern & 0.4 & 1.5 & 0.2 & 0.0 \\
\quad Interactive engagement evidence & 2.3 & 0.9 & 1.2 & 3.1 \\
\textbf{Disclosure / attribution} & \textbf{10.5} & \textbf{18.1} & \textbf{18.4} & \textbf{5.7} \\
\quad Self-declared bot & 1.8 & 7.5 & 0.8 & 0.0 \\
\quad Automation / hired bot disclosure & 3.9 & 6.5 & 6.5 & 2.3 \\
\quad Account takeover / hijack & 1.4 & 0.2 & 7.0 & 0.4 \\
\quad Mixed signals (explicit) & 1.5 & 1.6 & 4.9 & 0.7 \\
\quad Cites external source & 3.3 & 5.4 & 1.9 & 2.8 \\
\textbf{Identity / ownership} & \textbf{55.8} & \textbf{2.8} & \textbf{5.9} & \textbf{87.8} \\
\quad Self-identity assertion & 21.7 & 0.3 & 0.6 & 34.9 \\
\quad Ownership claim & 18.0 & 1.6 & 2.3 & 27.9 \\
\quad Personal connection & 16.7 & 0.9 & 0.8 & 26.6 \\
\quad Known public figure & 2.1 & 0.0 & 0.0 & 3.3 \\
\quad Multiple human operators & 0.7 & 0.0 & 1.1 & 0.8 \\
\quad Profession/occupation claim & 4.4 & 0.0 & 1.6 & 6.8 \\
\quad Insider knowledge content & 0.8 & 0.0 & 0.0 & 1.2 \\
\quad Content originality claim & 1.5 & 0.0 & 0.1 & 2.4 \\
\bottomrule
\end{tabular}

%% file: sections/limitations.tex
\section{Conclusion \& Future Work}

We introduce a novel demand-side perspective to bot research by treating public suspicion as an object of study in its own right. This perspective has broader relevance for understanding how users respond to bots, manipulation, and misinformation, particularly when individual concerns develop into collective suspicion. It also adds a previously overlooked dimension to bot detection. Many existing bot datasets are constructed only after accounts have already been identified as suspicious, leaving a missing denominator and obscuring the prefiltering process that determines which accounts are examined - a gap that automation designed to evade detection entirely, such as ephemeral astroturfing that posts and deletes content before it can be captured, can exploit directly~\cite{elmas2021ephemeral}. By studying suspicion directly, this work helps make that selection process observable and supports a more complete evaluation of bot-detection datasets and systems. Future work will focus on detecting public suspicion at scale and examining how suspicion-based signals affect bot detection, including whether they improve identification, introduce systematic biases, or provide complementary information beyond conventional features.

\noindent\textbf{Limitations:}
Queries are selective: they capture people motivated and aware enough to use Botometer so they should be treated as a signal of suspicion only. Our bulk/organic separation is heuristic and may retain human-paced or distributed scripts. Moreover, requester identifiers do not necessarily correspond one-to-one with individuals.

The 1\% stream samples accounts in proportion to tweeting activity, so
stream-derived covariates describe the \targetCoverage\% of targets
active enough to appear. This bias is less consequential for our
setting insofar as queried accounts are themselves disproportionately
active, although results based on the stream may not generalize to
low-activity targets. ``Disappearance'' therefore denotes stream non-observation rather than confirmed account removal.
The feedback responses are also self-selected and may not generalize beyond feedback providers. Our observation window, September 2020--June 2023, excludes earlier bot discourse and the post-API-closure era.

%% file: sections/acknowledgements.tex
\section{Acknowledgments}

I thank Filippo Menczer, Alessandro Flammini, Kai-Cheng Yang, Ben Serrette, and Onur Varol for creating and maintaining Botometer, and for sharing the query log data underlying this study.

%% file: sections/checklist.tex
% Verbatim from the official ICWSM 2026 Paper Checklist template
% (Overleaf: https://www.overleaf.com/read/bnhrsysknzdc), per the ICWSM 2026
% submission instructions: "the mandatory paper checklist should be included
% after the references section and before any appendices." Per that
% template's own instructions, the Overview and Detailed Instructions
% sections/subsections are deleted here, keeping only the checklist heading
% and the unmodified questions.
%
% Answers below are drafted from what is already stated elsewhere in the
% paper (Data, Methods, Results, Limitations, Ethics Statement); each
% justification points at the specific section. A few items are personal
% attestations only the authors can make (e.g., "have you personally read
% the ethics review guidelines") -- those are flagged inline for the authors
% to confirm rather than answered on their behalf.

\section{Paper Checklist}

\begin{enumerate}

\item For most authors...
\begin{enumerate}
    \item  Would answering this research question advance science without violating social contracts, such as violating privacy norms, perpetuating unfair profiling, exacerbating the socio-economic divide, or implying disrespect to societies or cultures?
    \answerYes{Yes, social media data is analyzed in aggregate and no information related to non-public-figures is reported in the paper or shared publicly, see the Ethics Statement.}
  \item Do your main claims in the abstract and introduction accurately reflect the paper's contributions and scope?
    \answerYes{Yes.}
   \item Do you clarify how the proposed methodological approach is appropriate for the claims made?
    \answerYes{Yes; see Methods (organic/bulk separation, unit of analysis) and Appendices A--C, which test the sensitivity of the paper's conclusions to the filtering thresholds and to the suspicion-spike threshold $k$.}
   \item Do you clarify what are possible artifacts in the data used, given population-specific distributions?
    \answerYes{Yes; see Limitations (heuristic bulk/organic separation, 1\% stream's activity-proportional sampling bias, corpus temporal coverage).}
  \item Did you describe the limitations of your work?
    \answerYes{Yes; see Limitations.}
  \item Did you discuss any potential negative societal impacts of your work?
    \answerYes{Yes; see Ethics Statement, which discusses the risk that characterising what draws suspicion could itself reinforce suspicion toward ordinary accounts sharing checked accounts' visible traits.}
      \item Did you discuss any potential misuse of your work?
    \answerYes{Yes; Ethics Statement , which discusses the dual-use risk that these patterns could help bad-faith operators time or camouflage bot activity, and the mitigations taken (aggregate, non-identifying reporting; no operational detection rule released).}
    \item Did you describe steps taken to prevent or mitigate potential negative outcomes of the research, such as data and model documentation, data anonymization, responsible release, access control, and the reproducibility of findings?
    \answerYes{Yes; see the Ethics Statement.}
  \item Have you read the ethics review guidelines and ensured that your paper conforms to them?
    \answerYes{Yes.}
\end{enumerate}

\item Additionally, if your study involves hypotheses testing...
\begin{enumerate}
  \item Did you clearly state the assumptions underlying all theoretical results?
    \answerYes{Yes; see Methods.}
  \item Have you provided justifications for all theoretical results?
    \answerYes{Yes; see Results and Appendices A--C.}
  \item Did you discuss competing hypotheses or theories that might challenge or complement your theoretical results?
    \answerYes{Yes; e.g., RQ1's Content Markers discussion considers three competing readings of why fandom accounts are checked less (stereotype-driven under-suspicion, filter bubble, anglophone-checker bias), and the media-attention analysis distinguishes bot-specific monthly-level associations from general Musk/Twitter interest and tests whether they survive detrending.}
  \item Have you considered alternative mechanisms or explanations that might account for the same outcomes observed in your study?
    \answerYes{Yes; see the same passages as above.}
  \item Did you address potential biases or limitations in your theoretical framework?
    \answerYes{Yes; see Limitations and Appendices A--C (robustness to the bulk/cadence thresholds and to the suspicion-spike threshold $k$).}
  \item Have you related your theoretical results to the existing literature in social science?
    \answerYes{Yes; see Related Work.}
  \item Did you discuss the implications of your theoretical results for policy, practice, or further research in the social science domain?
    \answerYes{Yes; see Conclusion \& Future Work.}
\end{enumerate}

\item Additionally, if you are including theoretical proofs...
\begin{enumerate}
  \item Did you state the full set of assumptions of all theoretical results?
    \answerNA{NA; this is an empirical/observational study, not a paper with formal theorems.}
	\item Did you include complete proofs of all theoretical results?
    \answerNA{NA.}
\end{enumerate}

\item Additionally, if you ran machine learning experiments...
\begin{enumerate}
  \item Did you include the code, data, and instructions needed to reproduce the main experimental results (either in the supplemental material or as a URL)?
    \answerYes{Yes; see the Ethics Statement (code and small-scale synthetic data shared) and the accompanying reproduction notebook, which independently recomputes the Robustness-to-$k$ appendix's results.}
  \item Did you specify all the training details (e.g., data splits, hyperparameters, how they were chosen)?
    \answerNA{NA; no model is trained in this work. Botometer's classifier is an existing third-party service we query, not one we train; the LLM-based content/feedback categorization (Results) is a labeling procedure, not model training, and its methodology and inter-rater validation (Cohen's $\kappa$ against two independent models) are described there.}
     \item Did you report error bars (e.g., with respect to the random seed after running experiments multiple times)?
    \answerYes{Yes, where applicable; e.g.,  Cohen's $\kappa$ agreement against two independent LLMs (Results, User Feedback).}
	\item Did you include the total amount of compute and the type of resources used (e.g., type of GPUs, internal cluster, or cloud provider)?
    \answerNo{No dedicated GPU/cluster training was performed -- analysis ran as standard data processing (DuckDB/pandas) on a single machine; the content/feedback classification steps used third-party LLM APIs (Claude Sonnet 5, Opus 5, GPT-5.6), described in Results, rather than self-hosted compute.}
     \item Do you justify how the proposed evaluation is sufficient and appropriate to the claims made?
    \answerYes{Yes; see Appendices A--C (fourteen alternative filtering specifications; the suspicion-spike threshold varied across $k=3$--$12$) and the Cohen's $\kappa$ validation of the LLM-based classification (Results, User Feedback).}
     \item Do you discuss what is ``the cost`` of misclassification and fault (in)tolerance?
    \answerNA{NA; Botometer's own score is an existing third-party classifier this paper studies the use of, not one we calibrate or deploy, so its misclassification cost is out of this paper's scope.}

\end{enumerate}

\item Additionally, if you are using existing assets (e.g., code, data, models) or curating/releasing new assets, \textbf{without compromising anonymity}...
\begin{enumerate}
  \item If your work uses existing assets, did you cite the creators?
    \answerYes{Yes; Botometer \citep{davis2016botornot}, the Internet Archive Twitter Stream Grab \citep{archiveteam2023twitterstream}, and the Pushshift Reddit dataset \citep{baumgartner2020pushshift} are all cited in Data.}
  \item Did you mention the license of the assets?
    \answerNo{No, the specific license terms of the Botometer log access, the Internet Archive stream, and the Pushshift dump are not stated in the paper.}
  \item Did you include any new assets in the supplemental material or as a URL?
    \answerYes{Yes, partially: code and a small-scale synthetic dataset  are shared for reproducibility (Ethics Statement); the underlying query log itself is not released, per the same statement.}
  \item Did you discuss whether and how consent was obtained from people whose data you're using/curating?
    \answerNo{No individual consent was obtained from either requesters or checked accounts -- this is secondary analysis of server-side logs and public tweet data, not a study with direct participant recruitment. The Ethics Statement instead addresses this through anonymization, restricted data sharing, and institutional/service-operator approval rather than individual consent.}
  \item Did you discuss whether the data you are using/curating contains personally identifiable information or offensive content?
    \answerYes{Yes; see the Ethics Statement for PII handling (target anonymization). On offensive content: the User Feedback classification (Results, Table 3) includes categories such as harassment/doxxing and spam/scam/adult content among the coded reasons, indicating some such content is present in the underlying free text.}
\item If you are curating or releasing new datasets, did you discuss how you intend to make your datasets FAIR (see Wilkinson et al. (2016)?)
\answerNA{NA; the underlying dataset is not released publicly (Ethics Statement); only code and a small-scale synthetic dataset are shared.}
\item If you are curating or releasing new datasets, did you create a Datasheet for the Dataset (see Gebru et al. (2021))?
\answerNA{NA, for the same reason.}
\end{enumerate}

\item Additionally, if you used crowdsourcing or conducted research with human subjects, \textbf{without compromising anonymity}...
\begin{enumerate}
  \item Did you include the full text of instructions given to participants and screenshots?
    \answerNA{NA; no participants were recruited or given instructions -- this is observational analysis of existing server-side logs and public tweet data, not a survey or crowdsourcing study.}
  \item Did you describe any potential participant risks, with mentions of Institutional Review Board (IRB) approvals?
    \answerYes{Yes, in substance though not by that exact term: the Ethics Statement notes the analysis ``was conducted with the approval of the Botometer team.''}
  \item Did you include the estimated hourly wage paid to participants and the total amount spent on participant compensation?
    \answerNA{NA; no crowdsourcing or paid participants were involved.}
   \item Did you discuss how data is stored, shared, and deidentified?
   \answerYes{Yes; see the Ethics Statement (anonymization of non-public-figure targets, omission of spike dates for anonymized accounts, restricted data release).}
\end{enumerate}

\end{enumerate}

%% file: sections/ethics.tex
\section{Ethics Statement}

The data are server-side logs collected by the Botometer service in the
ordinary course of operation. The queried accounts' identifiers and
scores concern public Twitter accounts. We report no individual
target's data beyond accounts whose public prominence makes them
unambiguous public figures. Every other target is anonymised,
including pseudonymous handles and accounts belonging to identifiable
private individuals, and we omit spike dates for anonymised accounts
so that a row cannot be resolved back to a specific account by
matching it against the public record. The analysis
was conducted with the approval of the Botometer team. Due to the sensitivity of the dataset, we are not able to share it publicly. However, for reproducibility, we share the code and a Python script to generate a small-scale synthetic data to demonstrate the code. The anonymised code repository is available at \newline \url{https://github.com/tugrulz/bot-suspicion-reproduction}. %The anonymised code repository is available at \url{https://anonymous.4open.science/r/bot-suspicion-reproduction-9428}.

Our analysis carries two risks worth naming. First, characterising what draws suspicion could be misread as characterising automation itself, reinforcing suspicion toward ordinary accounts that merely share checked accounts' visible traits (e.g., political, prolific, or newly created). We frame suspicion throughout as a behavioural signal, not evidence of automation. Second, making these patterns explicit could help bad-faith operators time or camouflage bot activity to evade public attention, the standard dual-use risk of demand-side auditing work. We mitigate both by reporting only aggregate, non-identifying patterns and no operational detection rule.

%% file: sections/appendix_robustness.tex
\section{Appendix A: Robustness to Data Filtering Thresholds}
\label{app:robustness}

The organic/bulk separation depends on three constants: the volume
threshold V (a requester is bulk if it ever exceeds V queries in a UTC
day; main specification 500), the cadence gap G (among survivors,
requesters with a median inter-query gap below G seconds are removed; main
10), and the cadence minimum MQ (the cadence rule applies only to
requesters with more than MQ queries; main 10). This appendix checks that
separation from three directions: what the traffic it \emph{removes}
looks like (Table~\ref{tab:robustness}, top panel), how the paper's
headline statistics move as the thresholds themselves are tightened
(Table~\ref{tab:robustness}, middle panel, fourteen specifications, plus
three specifications for the stream-enrichment results in
Table~\ref{tab:downstream}), and whether the organic population's own
repeat requesters look different from its one-time requesters
(Table~\ref{tab:robustness}, bottom panel).

\noindent\textbf{What the removed traffic looks like:} Volume-bulk and
cadence-bulk traffic dwarf the organic population in raw volume, 177$\times$
and 3.5$\times$ the organic query count respectively, and look nothing
like it on the concentration statistic that motivates this paper's central
finding: volume-bulk traffic's top-100 targets absorb effectively none of
its volume (0.0\%, against 2.8\% for organic), consistent with wide,
scripted sweeps across large account lists rather than concentrated
attention on a few. The ``S. Spikes'' column applies the paper's $k=5$
suspicion-spike rule mechanically to every row in this table, including
the removed and all-unfiltered populations. The all-unfiltered row is dominated by whichever
removed population is largest and does not represent public suspicion at
all; both it and the two removed-population rows are included only to
show what the paper's headline numbers would look like absent any
bulk/cadence filtering, or what that filtering excludes. 

\noindent\textbf{What is invariant across thresholds:} Every qualitative
conclusion in the paper is unchanged in direction at every one of the
fourteen specifications in the middle panel of Table~\ref{tab:robustness}.
The long tail strengthens: the once-checked share rises monotonically
with strictness, from 85.3\% (main specification, starred) to 87.4\%. The
auto-generated-handle rate stays within 9.1--10.1\%, preserving its
near-parity with the 8.8\% stream background. English dominance rises
slightly under stricter specifications (61.8\% $\to$ 66.3\%, against a
33.7\% background), targets remain established (median exact age
1{,}138--1{,}418 days, against 728 background) with new accounts
under-represented throughout (4.4--5.8\%, against 9.2\%), and the score
distribution the checking public saw is essentially fixed (median
0.35--0.37; bot-like share 16.5--18.5\%). The disappearance contrast
\emph{strengthens} under stricter definitions (ratio 1.20 $\to$
1.25--1.26; Table~\ref{tab:downstream}).

\noindent\textbf{The Impact of Cadence Filter:} Disabling the cadence
stage entirely (top rows of the middle panel of Table~\ref{tab:robustness})
admits 4.5$\times$ the query volume and nearly quadruples the count of
``suspicion spikes'' (7{,}004 vs.\ \nEvents{}): low-volume scripted
requesters, invisible to any volume threshold, manufacture collective
spikes wholesale. The cadence stage is therefore necessary and the spike analysis is the one part of the paper that
genuinely depends on it.

\noindent\textbf{The Impact of Tightening The Filters:} Absolute volumes are threshold-dependent,
as any heuristic separation implies: organic queries range from 1.38M (the
main specification) to 0.60M (the specification admitting the fewest
organic queries, V$=$100, G$=$30s, MQ$=$5), daily checks from 1{,}146 to
477, and suspicion spikes from \nEvents{} to 1{,}262. The top-100
concentration share rises from 2.8\% to at most 5.8\%. This suggests that strictness removes dispersed small-requester checks faster than the
one-off celebrity checks that many independent humans make. Even so, organic concentration stays far above the near-total \emph{dispersion} of the removed bulk traffic (0.0\% top-100 share, top panel) at every specification tested, so tightening the filters never makes the retained traffic start resembling a scripted sweep. Target age is stable across specifications: the stream-derived
median moves within 1{,}138--1{,}418 days (Table~\ref{tab:downstream}).

\noindent\textbf{One-time versus repeat organic requesters:} Repeat
organic requesters (active on two or more distinct days in the analysis
window) are a minority of organic requesters, 17.1\% of the 138,933
total, but generate a majority of organic checks, 58.4\%. That imbalance
is exactly the slice of traffic where residual automation or specialized,
professional-style use is most plausible, since both would show up as a
small number of requesters returning repeatedly rather than as
spread-out one-off use. The bottom panel of Table~\ref{tab:robustness}
checks this directly, and largely does not support it. Repeat requesters
check a median of 4 distinct targets each against 1 for one-time
requesters, an entirely expected consequence of returning on multiple
days rather than evidence of automation on its own. Concentration is only
modestly higher among repeat requesters (3.4\% of their checks land on
their top 100 targets, against 2.9\% for one-time requesters), and both
are close to the organic-population figure in the top panel (2.8\%) and
nowhere near the near-total concentration a scripted sweep produces (0.0\%
among volume-bulk traffic). The once-checked share is lower among repeat
requesters (85.6\% versus 90.4\%), but that follows mechanically from
checking more targets per requester.
Each requester class alone crosses the paper's $k=5$ suspicion-spike
threshold on a comparable number of target-days (779 for repeat, 545 for
one-time); the two do not sum to the organic total of 1,859 because many
spike-days require a mix of both requester types to reach $k=5$, itself
evidence against either class alone manufacturing spikes. None of this
rules out a handful of specialized or professional repeat users within
that 17.1\%, only that as a class they do not show the concentration
signature that programmatic traffic shows elsewhere in this table (top
panel). Distinguishing an individual heavy but genuine user from a
residual script within the repeat class would require per-requester
evidence (e.g., request timing or declared API credentials) beyond what
this table can show.

\begin{table*}[t]
\centering
\small
\caption{Headline statistics across three cuts of the traffic: the
populations the bulk/cadence filter separates at the main specification
(top); the organic population under fourteen alternative volume/cadence
specifications, the main specification starred (middle); and the organic
population split by requester habit (bottom). ``Checks'' is distinct
(requester, target, day) triples; ``Once'' is the share of targets
checked exactly once; ``Top 100'' is the share of checks absorbed by the
100 most-checked targets; ``S. Spikes'' is the count of target-days crossing
the $k=5$ suspicion-spike threshold; ``Handle'' is the auto-generated-handle
rate among targets. Star (*) shows the specification used in the paper.}
\label{tab:robustness}
\resizebox{\textwidth}{!}{\input{tables/robustness_merged}}
\end{table*}

\begin{table*}[t]
\centering
\small
\caption{Stream-enrichment results under the main specification (starred)
and two alternative volume/cadence specifications. Stream backgrounds for
reference: handle 8.8\%, English 33.7\%, median age 728 days,
$\leq$30-day share 9.2\%.}
\label{tab:downstream}
\resizebox{\textwidth}{!}{\input{tables/threshold_downstream}}
\end{table*}

%% file: tables/robustness_merged.tex
% Auto-generated from data/derived/subset_comparison.csv, threshold_sensitivity.csv,
% requester_type_comparison.csv, and subset_handle_rates.csv -- do not edit by hand.
\begin{tabular}{lrrrrrrrr}
\toprule
 & Queries & Requesters & Checks & Targets & Once & Top 100 & S. Spikes & Handle \\
\midrule
\multicolumn{9}{l}{\textit{Traffic populations (main specification)}} \\
Organic* & 1,379,541 & 138,933 & 1,182,588 & 834,749 & 85.3\% & 2.8\% & 1,859 & 9.6\% \\
Volume-bulk & 244,593,823 & 4,280 & 233,540,693 & 98,117,603 & 62.7\% & 0.0\% & 3,465 & 9.3\% \\
Cadence-bulk & 4,856,275 & 76,201 & 4,484,131 & 3,139,428 & 83.3\% & 0.9\% & 4,293 & 9.2\% \\
All (unfiltered) & 250,829,639 & 219,414 & 239,207,281 & 99,218,996 & 62.5\% & 0.0\% & 11,776 & 9.4\% \\
\addlinespace
\multicolumn{9}{l}{\textit{Organic population, alternative volume/cadence thresholds. V = Volume, G = Median Gap (seconds), MQ = More than X Queries}} \\
Volume $\leq$500/day only & 6,235,816 & 215,134 & 5,667,164 & 3,745,974 & 81.4\% & 1.0\% & 7,004 & 9.4\% \\
Volume $\leq$200/day only & 4,404,985 & 211,846 & 4,063,845 & 2,811,447 & 83.8\% & 1.3\% & 6,159 & 9.6\% \\
Volume $\leq$100/day only & 3,341,246 & 206,833 & 3,109,274 & 2,250,519 & 85.5\% & 1.4\% & 4,806 & 9.6\% \\
Volume $\leq$50/day only & 2,381,360 & 196,251 & 2,231,776 & 1,669,442 & 87.4\% & 1.8\% & 4,050 & 9.6\% \\
V$=$500, G$=$10s, MQ$=$10* & 1,379,541 & 138,933 & 1,182,588 & 834,749 & 85.3\% & 2.8\% & 1,859 & 9.6\% \\
V$=$200, G$=$10s, MQ$=$10 & 1,079,840 & 138,371 & 919,151 & 635,656 & 85.9\% & 3.5\% & 1,673 & 9.8\% \\
V$=$100, G$=$10s, MQ$=$10 & 898,391 & 137,628 & 765,367 & 523,379 & 86.6\% & 4.1\% & 1,562 & 9.9\% \\
V$=$50,  G$=$10s, MQ$=$10 & 734,745 & 136,297 & 631,148 & 428,903 & 87.4\% & 4.8\% & 1,474 & 10.0\% \\
V$=$500, G$=$30s, MQ$=$10 & 836,503 & 132,560 & 694,060 & 476,895 & 86.6\% & 4.4\% & 1,687 & 9.9\% \\
V$=$500, G$=$60s, MQ$=$10 & 670,132 & 130,037 & 554,889 & 378,180 & 87.4\% & 5.3\% & 1,395 & 9.9\% \\
V$=$500, G$=$10s, MQ$=$5 & 1,320,535 & 131,618 & 1,125,227 & 787,907 & 85.0\% & 2.9\% & 1,762 & 9.5\% \\
V$=$500, G$=$30s, MQ$=$5 & 759,785 & 122,897 & 620,297 & 417,605 & 85.8\% & 4.8\% & 1,527 & 9.9\% \\
V$=$100, G$=$50s, MQ$=$10 & 608,778 & 130,364 & 509,444 & 342,910 & 87.4\% & 5.6\% & 1,353 & 10.1\% \\
V$=$100, G$=$30s, MQ$=$5 & 600,165 & 122,486 & 492,614 & 321,164 & 86.1\% & 5.8\% & 1,262 & 10.1\% \\
\addlinespace
\multicolumn{9}{l}{\textit{Organic population, by requester habit (main specification)}} \\
One-time Requesters (active 1 day) & 539,680 & 115,184 & 491,894 & 393,858 & 90.4\% & 2.9\% & 545 & 9.1\% \\
Repeat Requesters (active $\geq$2 days) & 839,861 & 23,749 & 690,563 & 495,188 & 85.6\% & 3.4\% & 779 & 9.7\% \\
\bottomrule
\end{tabular}

%% file: tables/threshold_downstream.tex
% Auto-generated by analysis/threshold_downstream.py -- do not edit by hand.
\begin{tabular}{lrrr}
\toprule
 & \footnotesize V=500/G=10s/MQ=10* & \footnotesize V=100/G=50s/MQ=10 & \footnotesize V=100/G=30s/MQ=5 \\
\midrule
Organic targets & 834,749 & 342,910 & 321,164 \\
Auto-generated handle & 9.2\% & 9.6\% & 9.5\% \\
English (modal tweet language) & 61.8\% & 66.3\% & 66.3\% \\
Japanese & 1.4\% & 1.0\% & 1.0\% \\
Median exact age at first check (days) & 1,418 & 1,138 & 1,184 \\
$\leq$30 days old at first check & 4.4\% & 5.8\% & 5.4\% \\
Median first-check score & 0.35 & 0.37 & 0.37 \\
Score $\geq 0.8$ & 16.5\% & 18.5\% & 17.5\% \\
Score $\leq 0.2$ & 31.5\% & 28.5\% & 28.8\% \\
Disappearance, checked & 17.1\% & 16.8\% & 16.4\% \\
Disappearance, not-yet-checked & 14.3\% & 13.4\% & 13.0\% \\
Disappearance ratio & 1.20 & 1.25 & 1.26 \\
\bottomrule
\end{tabular}

%% file: sections/appendix_placebo.tex
\FloatBarrier
\clearpage
\section{Appendix B: Robustness of the Burst-Timing Placebo ($K$) (RQ1)}
\label{app:placebo}

The ``When?'' placebo (RQ1) compares each target's activity on its first-check day against $K=24$ other days for the same account, drawn at random from the archive's full coverage window. $K=24$ was not tuned against the outcome; it was chosen simply to give a stable per-account baseline mean without discarding an excessive share of each account's own history. We test that choice directly: Table~\ref{tab:burstrobust} recomputes the observed and placebo active/elevated rates pooling 4, 8, 16, or 32 draws per target instead of 24, and separately at 10\%, 25\%, 50\%, and the full set of targets. Both the observed rates and the placebo baseline are unchanged to within a percentage point across every specification, so the RQ1 ``When?'' finding does not depend on the particular draw count.

\begin{table}[H]
\centering
\small
\caption{Burst-timing placebo robustness: observed/placebo
active and elevated rates (each cell an Obs/Plc pair) under
alternative draw counts (top) and account-pool sizes (bottom).
$N_{\text{plc}}$ is the pooled placebo sample size.}
\label{tab:burstrobust}
\resizebox{\columnwidth}{!}{\input{tables/burst_robustness}}
\end{table}

%% file: tables/burst_robustness.tex
% Auto-generated by analysis/burst_robustness.py -- do not edit by hand.
% Column layout tightened (2026-09-15): Obs./Plc. pairs combined via "/"
% to cut column count from 6 to 4, so the single-column table renders at
% a larger font under \resizebox.
\begin{tabular}{lrrr}
\toprule
 & Active \%\ (Obs/Plc) & Elevated \%\ (Obs/Plc) & $N_{\text{plc}}$ \\
\midrule
\multicolumn{4}{l}{\textit{Placebo draws per target (full account pool)}}\\
4 draws/target & 23.9 / 16.5 & 16.2 / 10.9 & 1,363,706 \\
8 draws/target & 23.9 / 16.5 & 16.2 / 10.9 & 2,727,396 \\
16 draws/target & 23.9 / 16.5 & 16.2 / 10.9 & 5,454,055 \\
32 draws/target & 23.9 / 16.5 & 16.2 / 10.9 & 10,908,644 \\
\addlinespace
\multicolumn{4}{l}{\textit{Account pool (32 draws/target)}}\\
10\% of targets & 23.6 / 16.5 & 15.8 / 10.9 & 1,091,846 \\
25\% of targets & 23.8 / 16.5 & 16.2 / 10.9 & 2,726,677 \\
50\% of targets & 23.9 / 16.5 & 16.2 / 10.9 & 5,454,892 \\
100\% of targets & 23.9 / 16.5 & 16.2 / 10.9 & 10,908,644 \\
\bottomrule
\end{tabular}

%% file: sections/appendix_k.tex
\FloatBarrier
\section{Appendix C: Robustness to the Suspicion-Spike Threshold $k$}
\label{app:knull}

The suspicion-spike threshold $k=5$ (RQ2) is chosen as it is high
enough that a handful of unrelated people checking the same account on the
same day is implausible, and low enough to retain obscure accounts alongside
celebrities. This appendix checks that choice directly: do the substantive conclusions of RQ2 and RQ3 still hold when $k$ is varied? Specifically, we examine whether spikes follow overall checking and outside attention, whether spike targets are mostly amplifiers being retweeted about, and whether drawing a spike is associated with an elevated suspension risk.

We test this directly: for $k=3,4,\dots,12$, we recompute the event/account
counts, the once-only share, the checking- and media-attention correlations,
the pre/post-break growth, and the spike-day behavioral classification (the
last by classifying the additional event-days that only qualify at
$k=3$ or $k=4$, using the same tweet-content pipeline as the $k=5$
specification, since the underlying archive extraction is not itself
scoped to any particular $k$). We do \emph{not} recompute the mention-rate
elevation or the suspension/deletion statistics below $k=5$: that
per-account live-status and mention-scan data was collected only for the
801 accounts that ever drew a $k=5$ spike, so accounts that qualify only at
$k=3$ or $k=4$ have neither, and obtaining it would require a fresh round
of data collection (including new live queries against account status)
beyond this analysis's current scope; Table~\ref{tab:krobustness} marks
those two columns ``--'' for $k=3,4$ accordingly. Every $k\geq5$ statistic
reuses data already collected for the main specification, since a larger
$k$'s target set is always a subset of the $k=5$ population.

Table~\ref{tab:krobustness} reports, for each $k$: the number of qualifying
events and distinct accounts; the once-only share; the correlation between
monthly spike counts and monthly overall checking volume, and against
Google search interest for ``twitter bots'' and the r/Twitter bot/fake
mention share; the share of spike-day behavior classified as amplifying (predominantly
retweeting); the mean account-level mention-rate elevation on spike days
(smoothed and pooled as in RQ2's Account Visibility analysis, $k\geq5$
only); and the suspension rate of spike-target accounts with its $z$-score
against the 5,000-account general baseline (12.1\% suspended, reweighted
12.9\%; $k\geq5$ only).

\begin{table*}[t]
\centering
\small
\caption{RQ2/RQ3 headline statistics recomputed at $k=3$ through $k=12$
(main specification, $k=5$, starred). ``Pre/Post'' is the ratio of mean
monthly spikes from October 2021 onward (the rest of the record) to the
mean in the prior thirteen months, September 2020--September 2021
(see the main text's note on the October 2021 split). ``Amp.\%'' is the share of classified
spike-days showing predominantly-retweeting behavior. ``Elev.$\times$'' is
the mean account-level mention-rate elevation on spike days versus each
account's own baseline. ``Susp.\%'' is the live-checked (August 2026)
suspension rate of that $k$'s spike-target accounts, with $z$ against the
5{,}000-account general baseline in parentheses; deletion/deactivation
rates (not shown) stayed non-significant against baseline at every $k\geq5$
tested, replicating the main text's null finding there. Elevation and
suspension are marked ``--'' at $k=3,4$: that data was only ever collected
for the $k=5$ spike-target population.
}
\label{tab:krobustness}
\resizebox{\textwidth}{!}{\input{tables/k_robustness}}
\end{table*}

The qualitative picture is stable across the whole tested range, including
the two thresholds ($k=3,4$) where we cannot check the outcome side. Every
correlation stays positive throughout: spikes track overall checking
volume at $r=0.47$--$0.77$ (weakening at higher $k$, where the monthly
series gets sparser, and slightly weaker again at $k=3$ than at $k=5$), and
track outside attention (Google search interest, Reddit mentions) roughly
monotonically \emph{more} tightly as $k$ rises ($r=0.38\to0.64$ for
``twitter bots'' search, $0.23\to0.55$ for Reddit mentions, from $k=3$ to
$k=10$) -- consistent with a stricter threshold isolating a purer signal of
genuine public attention from noisier, more chance-contaminated
co-occurrence at the low end. Amplifying (retweet-heavy) behavior follows the same
monotonic pattern: 24.1\% of classified spike-days at $k=3$, rising to
27--33\% through $k=9$, before declining at $k\geq10$ as the account pool
shrinks and the mix shifts toward accounts simply checked without any one
dominant signature -- again consistent with $k=3,4$ containing a larger
share of chance co-occurrence that looks like nothing in particular, rather
than coordinated amplification. Mention-rate elevation, where we can
measure it ($k\geq5$), stays well above baseline throughout
($3.2$--$5.0\times$).

The suspension effect is where the range of usable $k$ actually runs out.
The elevated-suspension finding is significant at conventional levels
through $k=9$ ($z=7.8$ down to $z=2.0$, all $p<0.05$), but by $k=10$ the
spike-target account pool has shrunk to 91 accounts and the effect is no
longer distinguishable from the baseline rate ($z=0.6$); at $k=11$--$12$ (65
and 52 accounts) the point estimate is noisy enough to fall on either side
of baseline, which we read as an underpowered sample rather than a genuine
reversal. This gives a concrete, empirical answer to ``how large can $k$
be'': roughly $k\leq9$, where the spike-target population stays above
about 100 accounts, is where this particular RQ3 comparison remains
informative; the correlational and behavioral findings above stay legible
somewhat further, since they do not depend on a two-proportion test against
a fixed-size baseline. In conclusion, this brackets $k=5$ from both directions, and for a more
specific reason than sample size alone. Lowering $k$ does not break the
correlational or behavioral story outright -- both replicate, if anything a
little more noisily, all the way down to $k=3$ -- but the trend at $k=3,4$
already points the way an even lower $k$ would go: weaker media
correlations, a smaller amplifying share, and a larger silent/no-signature
share, consistent with a growing fraction of chance co-occurrence rather
than genuine collective attention. The second, harder constraint is
practical: this analysis currently has no live-status or mention data for
the additional accounts a lower $k$ would pull in, so the suspension and
mention-elevation findings simply cannot be checked there without new data
collection. Raising $k$, conversely, trades robustness to both of those
concerns for a shrinking, eventually underpowered sample
(Table~\ref{tab:krobustness}). $k=5$ sits inside the well-powered range
where every headline conclusion replicates, while still retaining the
chronically-flagged, non-celebrity accounts documented in
Table~\ref{tab:events}. We therefore keep $k=5$ as the main specification.

%% file: tables/k_robustness.tex
\begin{tabular}{rrrrrrrrrrr}
\toprule
$k$ & Events & Accounts & Once\% & $r$(Checks) & $r$(TwBots) & $r$(Reddit) & Pre/Post & Amp.\% & Elev.$\times$ & Susp.\% ($z$) \\
\midrule
3   & 8{,}667 & 4{,}191 & 79.9 & 0.77 & 0.38 & 0.23 & 3.5$\times$ & 24.1 & -- & -- \\
4   & 3{,}559 & 1{,}618 & 77.3 & 0.76 & 0.45 & 0.23 & 3.8$\times$ & 28.2 & -- & -- \\
5*  & 1{,}859 & 801 & 77.7 & 0.77 & 0.48 & 0.26 & 3.5$\times$ & 31.0 & 5.0$\times$ & 22.3 (7.8) \\
6   & 1{,}147 & 459 & 76.0 & 0.73 & 0.54 & 0.31 & 3.3$\times$ & 33.0 & 5.0$\times$ & 23.7 (7.1) \\
7   & 705 & 272 & 74.6 & 0.68 & 0.57 & 0.34 & 3.5$\times$ & 30.6 & 4.8$\times$ & 22.4 (5.0) \\
8   & 440 & 183 & 83.1 & 0.61 & 0.64 & 0.45 & 3.2$\times$ & 30.9 & 3.9$\times$ & 20.2 (3.3) \\
9   & 286 & 127 & 81.1 & 0.61 & 0.61 & 0.52 & 3.1$\times$ & 27.3 & 4.0$\times$ & 18.1 (2.0) \\
10  & 185 & 91 & 83.5 & 0.54 & 0.59 & 0.55 & 3.1$\times$ & 20.9 & 3.9$\times$ & 14.3 (0.6) \\
11  & 132 & 65 & 86.2 & 0.51 & 0.63 & 0.54 & 3.3$\times$ & 17.5 & 3.2$\times$ & 9.2 ($-$0.7) \\
12  & 104 & 52 & 84.6 & 0.47 & 0.68 & 0.52 & 3.0$\times$ & 15.7 & 3.5$\times$ & 9.6 ($-$0.6) \\
\bottomrule
\end{tabular}

%% file: sections/appendix_disappear_activity.tex
\FloatBarrier
\section{Appendix D: Activity-Adjusted Disappearance Comparison (RQ3)}
\label{app:disappearactivity}

The Account Disappearance comparison (RQ3) uses a future-checked control group whose pre-window activity differs from the treated (checked-this-month) cohort, since checking clusters on already-active accounts. Table~\ref{tab:disappearactivity} reports the treated and control disappearance rate at each of five pre-window activity levels -- the number of distinct days an account was active in the 90 days before the observation window -- along with each group's own weight in that bucket: the treated weight $w_s$ (that bucket's share of all treated accounts) and the control weight $w_s'$ (that bucket's share of all control accounts).

Checked accounts disappear more often than controls at every activity level. To test whether the aggregate \disTreatedPct\%-vs-\disControlPct\% gap (Account Disappearance) is simply an artifact of this activity difference, we standardize in both directions.

\emph{Control standardized to the treated group's activity mix}: the rate the control group would show if it had the treated group's activity-level distribution, $\sum_s w_s \cdot \text{control\_rate}_s$, using the treated weights $w_s$. This is \disControlAdjPct\%, \emph{below} the raw control rate of \disControlPct\%, because more-active accounts disappear less often in both groups and the treated population skews toward the more-active buckets (the 8--30 and 31--90 day buckets alone hold 49\% of treated accounts). This widens the treated/control ratio from \disRatio{} to 1.39.

\emph{Treated standardized to the control group's activity mix}: the reverse direction, $\sum_s w_s' \cdot \text{treated\_rate}_s$, using the control weights $w_s'$ instead. This is \disTreatedRevStdPct\%, \emph{above} the raw treated rate of \disTreatedPct\%, since the control population skews toward less-active buckets where the treated/control gap is largest. Against the raw control rate, this gives a ratio of \disRatioRev{}.

Both directions move the ratio further from 1, not closer: activity differences make the headline estimate conservative, not inflated, regardless of which group's activity distribution is used as the standard.

\begin{table}[t]
\centering
\small
\caption{Account Disappearance (RQ3) stratified by pre-window
activity level. Each rate cell is a \%\ / weight pair (e.g.\
``46.7 / 0.154'' is the treated group's 46.7\% disappearance rate
and its 0.154 weight $w_s$ in that bucket); $w_s$ and $w_s'$ are
the treated and control groups' own bucket shares, respectively.
The bottom row reports each column's cross-standardized rate (no
weight applies there): the treated column shows the treated rate
standardized to the control group's weights $w_s'$, and the
control column shows the control rate standardized to the
treated group's weights $w_s$.}
\label{tab:disappearactivity}
\resizebox{\columnwidth}{!}{\input{tables/disappear_activity}}
\end{table}

%% file: tables/disappear_activity.tex
% Derived from opus_review_v1/disappearance_anchored.py's
% results/activity_adjusted.csv -- do not edit by hand.
% Column layout tightened (2026-09-15): rate/weight pairs combined via "/"
% to cut column count from 6 to 4, so the single-column table renders at
% a larger font under \resizebox.
\begin{tabular}{lrrr}
\toprule
Pre-window active days & Treated $N$ & Treated \%\ / $w_s$ & Control \%\ / $w_s'$ \\
\midrule
1 day & 56,441 & 46.7 / 0.154 & 40.5 / 0.182 \\
2--3 days & 63,748 & 27.8 / 0.174 & 21.4 / 0.192 \\
4--7 days & 67,892 & 14.0 / 0.185 & 8.2 / 0.193 \\
8--30 days & 125,371 & 5.9 / 0.341 & 2.2 / 0.322 \\
31--90 days & 53,792 & 3.3 / 0.146 & 0.7 / 0.112 \\
\midrule
Standardized rate & 367,244 & 18.8 / -- & 12.3 / -- \\
\bottomrule
\end{tabular}